\documentclass[fleqn,usenatbib]{mnras}

\usepackage[T1]{fontenc}

\DeclareRobustCommand{\VAN}[3]{#2}
\let\VANthebibliography\thebibliography
\def\thebibliography{\DeclareRobustCommand{\VAN}[3]{##3}\VANthebibliography}

\usepackage{graphicx}	
\usepackage{amsmath}	
\usepackage{tikz}
\usepackage{CJK}
\usepackage{multirow}
\usepackage{algorithmicx}
\usepackage{algpseudocode}
\usepackage[caption=false]{subfig}
\usepackage{amssymb,amstext}
\usepackage{enumitem}

\usepackage{newtxtext,newtxmath}

\newcommand{\half}{\tfrac{1}{2}}
\newcommand{\p}{\partial}
\newcommand{\pc}{\mbox{\,pc}}
\newcommand{\kpc}{\mbox{\,kpc}}
\newcommand{\kms}{\mbox{\,km s}^{-1}}
\newcommand{\Myr}{\mbox{\,Myr}}
\newcommand{\Gyr}{\mbox{\,Gyr}}

\title[Origin and fate of the Gaia snail]{The origin and fate of the Gaia phase-space snail}

\author[Tremaine, Frankel and Bovy]{
Scott Tremaine,$^{1,2}$\thanks{E-mail: tremaine@ias.edu}
Neige Frankel,$^{1,3}$
and Jo Bovy$^3$
\\
$^{1}$Canadian Institute for Theoretical Astrophysics, University of Toronto, 60 St. George Street, Toronto, ON M5S 3H8, Canada\\
$^{2}$School of Natural Sciences, Institute for Advanced Study, Princeton, NJ 08540, USA\\
$^{3}$David A. Dunlap Department of Astronomy and Astrophysics, University of Toronto, 50 St. George Street, Toronto, ON M5S 3H4, Canada
}

\date{Accepted XXX. Received YYY; in original form ZZZ}

\pubyear{2022}

\begin{document}
\label{firstpage}
\pagerange{\pageref{firstpage}--\pageref{lastpage}}
\maketitle

\begin{abstract}

The Gaia snail is a spiral feature in the distribution of solar-neighbourhood stars in position and velocity normal to the Galactic midplane. The snail probably arises from phase mixing of gravitational disturbances that perturbed the disc in the distant past. The most common hypothesis is that the strongest disturbance resulted from a passage of the Sagittarius dwarf galaxy close to the solar neighbourhood. In this paper we investigate the alternative hypothesis that the snail is created by many small disturbances rather than one large one, that is, by Gaussian noise in the gravitational potential. Probably most of this noise is due to substructures in the dark-matter halo. We show that this hypothesis naturally reproduces most of the properties of the snail. In particular it predicts correctly, with no free parameters, that the apparent age of the snail will be $\sim0.5\Gyr$. An important ingredient of this model is that any snail-like feature in the solar neighbourhood, whatever its cause, is erased by scattering from giant molecular clouds or other small-scale structure on a time-scale $\lesssim 1\Gyr$.

\end{abstract}


\section{Introduction}

\label{sec:intro}

\begin{figure}
    \includegraphics[width=\columnwidth]{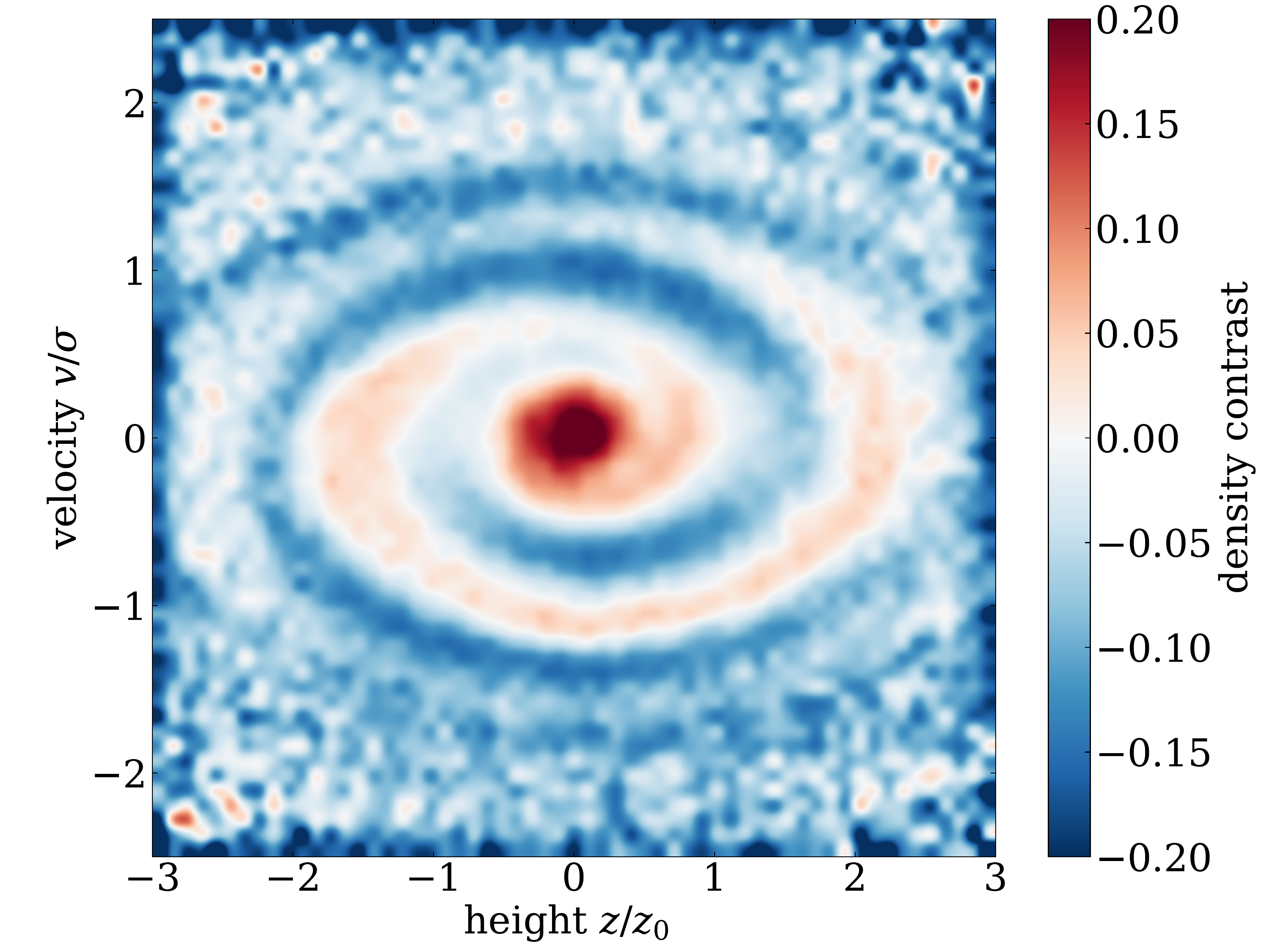}
    \caption{Fractional density contrast relative to a smooth distribution in the vertical or $z$--$v$ phase-space plane within $1\kpc$ of the Sun, revealing the snail in the Gaia DR3 data set \citep{gaiadr3}. Here we normalized the positions and velocities by $z_0=300\pc$ and $\sigma = 30\kms$. Adapted from \citet{fra22}.}
    \label{fig:snail}
\end{figure}

The Gaia snail or spiral \citep{antoja2018} is a prominent spiral feature in the vertical phase space of the solar neighbourhood. The snail is shown in Fig.\ \ref{fig:snail}, where the vertical phase-space coordinates are $(z,v)$, the height and velocity of stars in the direction normal to the Galactic midplane. A similar pattern is seen in plots of other quantities, such as the mean velocity components, as functions of $z$ and $v$. The snail is a dramatic illustration of the well-known fact that the Galactic disc is only approximately in a steady state \citep[e.g.][]{tremaine93}.

The snail is almost certainly an example of phase mixing \citep{arnold68,tremaine99}, which begins when some event or events produce a density enhancement or deficit in a region of phase space. The density perturbation then propagates through phase space according to the collisionless Boltzmann equation, becoming more and more distorted. In particular, in a typical galactic disc potential orbits with smaller excursions from the $z=0$ plane have shorter vertical periods, so a generic perturbation is sheared into a spiral that becomes more and more tightly wound with time. The spiral turns counter-clockwise with increasing size if the $(z,v=\dot z)$ phase space is a right-handed coordinate system as in Fig.\  \ref{fig:snail}.   

The most popular explanation for the event that gave birth to the snail is an encounter of the comoving solar neighbourhood with a satellite galaxy. Most satellites excite only negligible perturbations within the disc, because either their masses are too small or the encounters are too distant\footnote{Distant encounters cannot efficiently excite features in the vertical phase space because the encounter time -- roughly the impact parameter divided by the velocity of the perturber relative to the solar neighbourhood -- is longer than the vertical orbital period, so the orbits are adiabatically invariant.}. The strongest perturbations from a known source -- by an order of magnitude or more -- are due to the Sagittarius dwarf galaxy \citep{banik22}. 

There are some difficulties with a model in which an encounter with Sagittarius is the main stimulus for the snail. (i) In an extensive set of $N$-body simulations, \cite{bbh22} find that they are unable to match the observed amplitude of the snail unless the mass of the Sagittarius dwarf is much larger than the best current estimates, and that whatever the mass they are unable to find a detailed match to the shape of the snail. (ii) The estimated time of the encounter, derived by fitting the shape of the snail, varies by a factor of two or more, depending mainly on the area of the Galactic disc that is included in the sample\footnote{\label{foot:one} Estimates of the age of the encounter include 0.3--$0.9\Gyr$ \citep{antoja2018}, 0.5--$0.8\Myr$ \citep{laporte19}, $\sim0.54\Gyr$ \citep{lw21}, 0.3--$1.3\Gyr$ \citep{widmark22}, and 0.2--$0.6\Gyr$ \citep{fra22}.}. (iii) Other aspects of the snail morphology, in particular the transition from a one-armed spiral in the solar neighbourhood to a two-armed spiral in the inner Galaxy, suggest that at least two events have excited spirals in the disc \citep{hunt22}. (iv) The phase-space residuals between the data and models based on a single encounter appear to be dominated by coherent structures rather than random noise \citep{fra22}.

In this paper we argue that the origin and fate of the Gaia snail and similar features depend on the competing effects of large-scale and small-scale perturbations to the vertical phase-space distribution of stars. In this oversimplified picture, `large-scale' perturbations are those due to mass concentrations such as spiral structure, satellite galaxies, and dark-matter substructure that affect most or all of the stars in the solar neighbourhood. In contrast, `small-scale' perturbations are those that affect only a small fraction of the stars in the sample. A classic example of a small-scale perturbation is two-body relaxation due to gravitational encounters between individual stars, although this process is far too slow to affect nearby stars within the age of the universe. A much faster small-scale process is relaxation due to gravitational encounters with giant molecular clouds, which is likely to be the principal mechanism that drives the thickening of the stellar disc with age, at least in the solar neighbourhood \citep{carl1987,jb1990,ikm1993,sell2014,aumer2016,mackereth19}. The density of giant molecular clouds decreases sharply beyond the solar radius, and in this region small-scale perturbations are likely to be weaker and snails are likely to live longer, but our focus here is on the solar neighbourhood.  

\subsection{Dynamics of the snail}

\label{sec:snaildyn}

We now sketch the evolution of the snail using quantitative but approximate arguments. Let $t=0$ denote the time of birth of the Galactic disc and let $T$ denote the present. Suppose that a large-scale perturbation occurs at time $t_1$, $0<t_1<T$, and that the typical vertical orbital period in the solar neighbourhood is $P$ ($\sim 100\Myr$). The spiral produced by this perturbation will wind up about $w=\Gamma(T-t_1)/P$ times by the present time; here $\Gamma$ is a dimensionless factor depending on the relation between period and vertical amplitude of a stellar orbit in the disc (see eq.\ \ref{eq:fdef} for a precise definition), which we take to be $\Gamma\simeq 0.2$ (this choice is justified in \S\ref{sec:shear}). If the typical vertical velocity in the solar neighbourhood is $v$ then the velocity separation between peaks and troughs of the spiral will be about $\Delta v_\mathrm{lg}\sim\half v/w$. On the other hand, over this time interval small-scale perturbations cause the stellar velocities to diffuse by an amount $\Delta v_\mathrm{sm}^2\sim D_\mathrm{sm}(T-t_1)=D_\mathrm{sm}Pw/\Gamma$ where $D_\mathrm{sm}$ is the diffusion coefficient associated with these  perturbations. The snail will be erased when $\Delta v_\mathrm{sm} \gtrsim \Delta v_\mathrm{lg}$ or $w^3\gtrsim v^2\Gamma/(4D_\mathrm{sm}P)$. 

The diffusion coefficient $D_\mathrm{sm}$ can be estimated by assuming that the current vertical velocity dispersion $v$ has been excited mainly by small-scale perturbations. Then $v^2\sim D_\mathrm{sm}T$ so the snail disappears when its age $T-t_1$ exceeds $t_\mathrm{max}\simeq k\Gamma^{-2/3}P^{2/3}T^{1/3}$ with $k=2^{-2/3}=0.63$. An accurate derivation, described in the Appendix, shows that for stars at the mean action the amplitude of the snail decays as $\exp[-(T-t_1)^3/t_\mathrm{max}^3]$ with $k=3^{1/3}(2\pi)^{-2/3}=0.424$. Taking $P\sim 100\Myr$, $T\sim 10\Gyr$ and $\Gamma\simeq 0.2$ we find $t_\mathrm{max}\simeq 0.6\Gyr$ so no visible snail should be much older than this. See \S\ref{sec:wash} for a simulation of this process.

These arguments suggest a novel model for the origin of the snail. Suppose that the solar neighbourhood is subjected to a large number of weak large-scale perturbations. Let $(J,\theta)$ be action-angle variables for the vertical motion in the disc (eqs.\ \ref{eq:aa}) and suppose that a perturbation at time $t_i$ changes the distribution function by an amount $\Delta f_i(J,\theta)$. For simplicity we consider only the $m=1$ Fourier component in $\theta$, so we can write $\Delta f_i(J,\theta)=\Delta f_i(J)\cos[\theta-\theta_i(J)]$. For simplicity we shall assume that $\theta_i(J)$ is independent of $J$. At the present time $T$ this perturbation will have evolved to 
\begin{equation}
\Delta f_i(J)\cos[\theta-\Omega(J)(T-t_i)-\theta_i],
\label{eq:wrap}
\end{equation}
where $\Omega(J)$ is the frequency (eq.\ \ref{eq:omega}). The total perturbation to the present-day disc is then $\Delta f(J,\theta)=\sum_i \Delta f_i(J)\cos[\theta-\Omega(J)(T-t_i)-\theta_i]$ where the sum is over all of the perturbation events. We now calculate the average of the correlation function $\Delta f(J,\theta)\Delta f(J',\theta')$ over an ensemble of these histories, denoting that ensemble average by $\langle\cdot\rangle$, and assume that the phases $\theta_i$ are uncorrelated. Then $\langle\cos\theta_i\cos\theta_j\rangle=\langle\sin\theta_i\sin\theta_j\rangle=\half\delta_{ij}$ and  $\langle\sin\theta_i\cos\theta_j\rangle=0$, so the ensemble average of the correlation function in action-angle space is
\begin{align}
    &\langle \Delta f(J,\theta)\Delta f(J',\theta')\rangle \\&=\half\langle \Delta f(J)\Delta f(J')\rangle\sum_i \cos[\theta-\theta'-(\Omega-\Omega')(T-t_i)]\nonumber
\end{align}
where $\Omega'\equiv\Omega(J')$. To simplify the notation we have replaced $\langle \Delta f_i(J)\Delta f_i(J')\rangle$ by  $\langle \Delta f(J)\Delta f(J')\rangle$ since the ensemble average is independent of the event number $i$. 

We now assume, following the arguments earlier in this subsection, that the correlations are erased when the large-scale perturbations are older than $t_\mathrm{max}$. If there are many events occurring randomly at a uniform rate $r$, we can replace the sum by a time integral to obtain
\begin{align}
    &\langle \Delta f(J,\theta)\Delta f(J',\theta')\rangle\nonumber \\
    &=\half r\langle \Delta f(J)\Delta f(J')\rangle\int_0^{t_\mathrm{max}}\mathrm{d}\tau \cos[\theta-\theta'-(\Omega-\Omega')\tau]\nonumber \\
    &=\half r\langle \Delta f(J)\Delta f(J')\rangle\nonumber \\
    &\quad\times \frac{\sin(\theta-\theta')-\sin[\theta-\theta'-(\Omega-\Omega')t_\mathrm{max}]}{\Omega-\Omega'}.
    \label{eq:xi}
\end{align}
In this model the evolution of the distribution function is described by a Gaussian process\footnote{\cite{nw22} use Gaussian processes in a different context, as an empirical model for the Galaxy's in-plane velocity field.}, and therefore is specified fully by the correlation function (\ref{eq:xi}). For comparison the correlation function of a band-limited Gaussian white-noise process is $\langle\rho(x)\rho(x')\rangle \propto \sin[k_\mathrm{max}(x-x')]/(x-x')$. 

The spiral nature of the correlation function in equation (\ref{eq:xi}) suggests that the Gaia snail could be the result of many weak, independent, large-scale perturbations to the solar neighbourhood rather than a single large-scale perturbation. In this model the  characteristic age of the spiral is related to $t_\mathrm{max}$ rather than to a specific encounter time. In the following section we develop simple simulations to explore this possibility.

\section{Model}

\subsection{Phase space and canonical variables}

\label{sec:aa}

We use a model based on slab symmetry, that is, we ignore variations in the stellar distribution function within the Galactic plane. Thus the phase space is specified by the `vertical' coordinate $z$, the distance of a star in the solar neighbourhood from the Galactic midplane, and the corresponding velocity $v=\dot z$. We assume that the gravitational potential is that of a self-gravitating isothermal slab \citep{spitzer42},
\begin{equation}
  \Phi(z)=2\sigma^2\log\left[\cosh(\half z/z_0)\right],
  \label{eq:eqpot}
\end{equation}
which is generated by the density distribution
\begin{equation}
\rho(z)=\rho_0\mbox{\,sech}^2(\half z/z_0).
\end{equation}
Here $z_0$ is the scale height of the density distribution when $|z|\gg z_0$ and $\sigma^2=8\pi G\rho_0z_0^2$. The corresponding distribution function is
\begin{equation}
  f(z,v)=\frac{\rho_0}{(2\pi\sigma^2)^{1/2}}\exp\left[-H(z,v)/\sigma^2\right]
  \label{eq:df}
\end{equation}
where $H(z,v)=\half v^2+\Phi(z)$ is the Hamiltonian or energy.  With this distribution function, the mean-square velocity at any height $z$ is $\langle v^2\rangle=\sigma^2$.

In the potential (\ref{eq:eqpot}) we can define action-angle variables,
\begin{equation}
J=\frac{1}{2\pi}\oint v\,dz, \quad \theta=\Omega(J)\int_0^z\frac{dz}{v},
\label{eq:aa}
\end{equation}
where the orbital frequency is 
\begin{equation}
\Omega(J)=2\pi\left(\oint\frac{dz}{v}\right)^{-1}.
\label{eq:omega}
\end{equation}
When the action is small, $J\ll\sigma z_0$, the orbit is confined to the region $|z|\ll z_0$, the potential $\Phi(z)\simeq \frac{1}{4}\sigma^2z^2/z_0^2=2\pi G\rho_0z^2$, and the frequency is  $\Omega=(4\pi G\rho_0)^{1/2}$, independent of $J$. For $J\gg \sigma z_0$ we have $\Omega(J)=[\pi^2\sigma^4/(12z_0^2)]^{1/3}J^{-1/3}$.

The Hamiltonian can also be written as a function of the action, $H(J)$, and then $\Omega(J)=dH(J)/dJ$. The mean action is
\begin{equation}
  \langle J\rangle =1.7974\,\sigma z_0.
  \label{eq:jbar}
  \end{equation}

The angle $\theta$ is ill-defined near $J=0$, so we introduce a new Cartesian canonical coordinate $q$ and momentum $p$ defined by\footnote{The new variables are derived from the mixed-variable generating function $S(\theta,p)=\half p^2\tan\theta$ using the relations $J=\p S/\p\theta$, $q=\p S/\p p$.}
\begin{equation}
  q=(2J)^{1/2}\sin\theta, \quad p=(2J)^{1/2}\cos\theta.
  \label{eq:pqjt}
\end{equation}

\subsection{Shear and winding}

\label{sec:shear}

Using these results we can estimate the rate of winding of a feature in phase space. Stars at a given angle $\theta_0$ evolve after time $t$ to the locus $\theta(J,t)=\theta_0+\Omega(J)t$. Over a small interval $\Delta J$ in the action, the angle varies as $\Delta\theta= (\mathrm{d}\Omega/\mathrm{d} J)t\Delta J$. If the spiral is tightly wound, its wavelength in action space is given by $|\Delta\theta|=2\pi$ or $\Delta J=(2\pi/t)|\mathrm{d}\Omega/\mathrm{d} J|^{-1}.$ The number of winds of the spiral is approximately $w\equiv J/\Delta J=\Gamma(J)t/P(J)$, in which we define the dimensionless shear
\begin{equation}
\Gamma(J)\equiv \left|\frac{\mathrm{d}\log\Omega(J)}{\mathrm{d}\log J}\right|
\label{eq:fdef}
\end{equation}
and $P(J)=2\pi/\Omega(J)$ is the orbital period. For a Kepler potential $\Gamma=3$; for a harmonic-oscillator potential $\Gamma=0$.

\begin{figure}
    \includegraphics[width=\columnwidth]{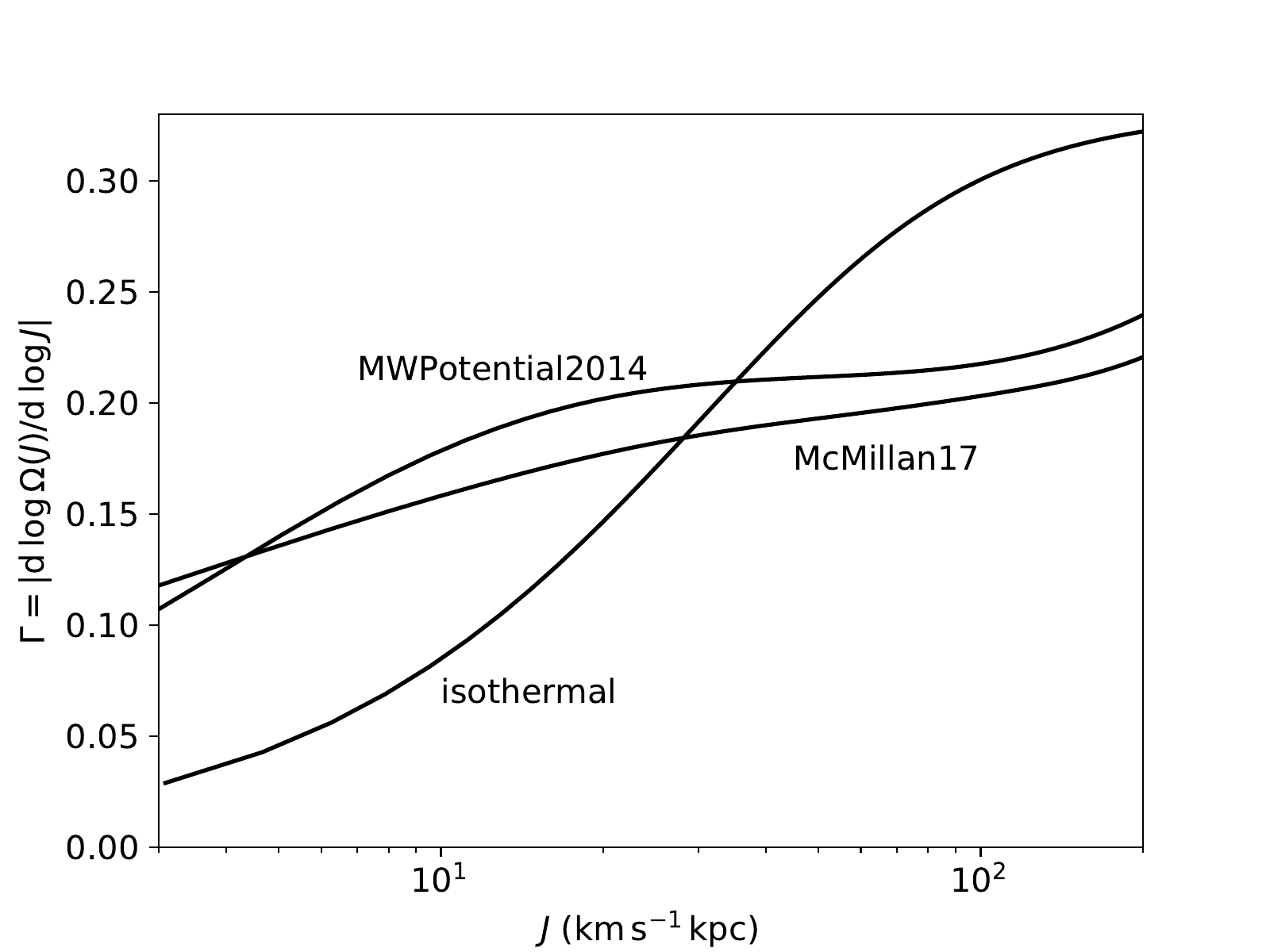}
    \caption{Shear $\Gamma=|\mathrm{d}\log\Omega/\mathrm{d}\log J|$ as a function of the action $J$. The shear is plotted for three gravitational potentials $\Phi(z)$: (i) the self-gravitating isothermal slab, equation (\ref{eq:eqpot}), with $z_0=300\pc$ and $\sigma=30\kms$; (b) the potential from \citet{mcmillan17}, evaluated at the solar radius; (c) the potential \textsc{MWPotential2014} from \citet{bovy15}, evaluated at the solar radius.}
    \label{fig:shear}
    \vspace{10pt}
\end{figure}

The shear for the potential (\ref{eq:eqpot}) is plotted in Fig.\  \ref{fig:shear} as the curve labelled `isothermal', assuming asymptotic scale height $z_0=300\pc$ and dispersion $\sigma=30\kms$. In this potential $\Gamma$ varies from nearly zero when $J\ll\sigma z_0$ ($\Phi(z)\propto z^2$) to $\frac{1}{3}$ for $J\gg \sigma z_0$ ($\Phi(z)\propto |z|$). At the mean action $J=\langle J\rangle=1.7974\,\sigma z_0=16.18\kms\kpc$, $\Gamma=0.125$.

We have also plotted the shear for two more realistic models of the vertical potential at the solar radius \citep{bovy15,mcmillan17}, as computed using the \textsc{galpy} \citep{bovy15} software package\footnote{\url{https://github.com/jobovy/galpy}~.}. These have shear curves that are flatter than the shear curve for the isothermal potential -- when the action is small the shears are larger because of the presence of mass components with small scale heights, and when the action is large the shears are smaller because of the potential from the central mass concentration and the dark halo. Over a factor of ten in action, from $J=10\kms\kpc$ to $J=100\kms\kpc$, the shear in these two potentials varies only over the limited range $\Gamma=0.16$--0.22. For the approximate arguments in \S\ref{sec:snaildyn} we adopt $\Gamma\simeq0.2$. 

\subsection{Small-scale kicks}

\label{sec:small}

For simplicity we assume that the small-scale perturbations are instantaneous kicks that can be modelled as an independent random walk in phase space for each star: thus the change in the coordinate $q$ over a small time $\Delta t$ is a Gaussian random variable with zero mean and variance $\langle(\Delta q)^2\rangle=D_q(q,p)\Delta t$, with a similar expression for the momentum $p$. We do not attempt to model the dynamics of the small-scale perturbations accurately; instead we simply set $D_q(q,p)=D_p(q,p)\equiv D_\mathrm{sm}$, where $D_\mathrm{sm}$ is a constant that we call the small-scale diffusion constant. Since $D_\mathrm{sm}$ is independent of $q$ and $p$, the variance in $q$ and $p$ due to small-scale kicks over an interval $\Delta t$ of \emph{any} length is just $\langle(\Delta q)^2\rangle=\langle(\Delta p)^2\rangle=D_\mathrm{sm}\Delta t$.

\subsection{Large-scale kicks}

We assume that the large-scale perturbations are instantaneous kicks $\Delta q_k(q,p), \Delta p_k(q,p)$ to the coordinates and momenta of all the stars, occurring at times $t_k$, where the kick times $t_k$ are distributed as a Poisson process with rate $r=K/T$ between $t=0$ (the formation time of the disc) and $t=T$ (the present time); thus the mean number of kicks is $K$, with $K\gg1$ by assumption. We do not attempt to model the dynamics of the kicks accurately; instead we assume that $\Delta q_k(q,p)$ and $\Delta p_k(q,p)$ are independent of $q$ and $p$ and are Gaussian random variables with zero mean and dispersion $\langle(\Delta q_k)^2\rangle=\langle(\Delta p_k)^2\rangle\equiv s^2$.

Since $K\gg1$ the total changes in the coordinates and momenta due to large-scale kicks are governed by the central limit theorem and therefore have Gaussian probability distributions with dispersions $rs^2T\equiv D_\mathrm{lg}T$ where $D_\mathrm{lg}$ is a constant that we call the large-scale diffusion constant. In other words the present distribution function depends on the rate $r$ and the mean-square kick $s^2$ only through the combination $rs^2=D_\mathrm{lg}$.

\subsection{Simulations}

We assume that the disc is initially cold, that is, all stars have action $J=0$ at time $t=0$. Then the mean value of the action at the present time $t=T$ is
\begin{equation}
  \langle J\rangle =\half\langle q^2+p^2\rangle= D_\mathrm{sm}T+s^2K=(D_\mathrm{sm}+D_\mathrm{lg})T.
\end{equation}
Introduce the notation
\begin{equation}
  D\equiv D_\mathrm{sm}+D_\mathrm{lg}, \quad D_\mathrm{sm}=(1-g)D,
  \quad D_\mathrm{lg}=gD;
  \label{eq:gdef}
\end{equation}
then $0<g<1$ is the fraction of the mean action in the disc that is due to large-scale kicks, while the remainder is due to small-scale kicks, and $\langle J\rangle=DT$. We may determine the diffusion coefficient $D$ by matching the mean action to the observed value at the present time (eq.\ \ref{eq:jbar}): in dimensionless units, 
\begin{equation}
  D=\frac{1.7974}{T}.
\end{equation}

The parameter $g$ is not well-determined from theory or observations. \cite{mj2016} examined vertical heating of the disc in $\Lambda$CDM cosmological simulations of halos similar to Milky Way and found that impacts with dark-matter substructure and satellite galaxies provided only 20--30\% of the heating rate observed in the solar neighbourhood. Similar simulations by \cite{grand2016} showed that substructure and satellite could produce a large fraction of the disc heating but only in rare cases. 

Our simple model of the excitation processes, both small-scale and large-scale, produces a distribution function of the form $f(J)\propto \exp(-J/\langle J\rangle)$ and predicts that the relation between the mean action $\langle J\rangle$ and the age $\tau$ of a coeval population of stars should have the form $\langle J\rangle\propto \tau$. Both of these results are consistent with observations \citep{tr19}. 

Two aspects of the model we have described are not strictly self-consistent: (i) We have assumed a fixed potential, equation (\ref{eq:eqpot}), while in fact the potential is changing because the scale height $z_0$ and dispersion $\sigma$ are growing with time. We do not believe this inconsistency affects our results, since we have argued in \S\ref{sec:snaildyn} that the properties of the snail are determined in the most recent Gyr, only ten percent of the Galaxy's age. (ii) As described in the preceding paragraph, our model produces a distribution function of the form $f(J)\propto \exp(-J/\langle J\rangle)$ whereas the distribution function we have assumed in equation (\ref{eq:df}) is $f(J)\propto\exp[-\beta H(J)]$. We do not believe that this inconsistency has a strong effect on our results: if we normalize the two distribution functions to have the same surface density and mean action $\langle J\rangle$ they differ by less than 12\% over the range $0<J<7.5\,\sigma z_0$. 

The pseudocode to carry out a simulation with $N$ stars and a rate $r=K/T$ of large-scale kicks is as follows. Here $x\leftarrow U(a,b)$ means `assign $x$ a random number uniformly distributed between $a$ and $b$' and $x,y\leftarrow N(\mu,\sigma^2)$ means `assign $x$ and $y$ random numbers from a Gaussian probability distribution with mean $\mu$ and dispersion $\sigma^2$'.

\begin{algorithmic}
  \State $(J_i,\theta_i)\leftarrow (0,0),\quad i=1,\ldots,N$
  \State $t\leftarrow 0$
  \State $k\leftarrow 0$
  \While{$t<T$} \Comment{generate kick times}
  \State $\Delta t\leftarrow -(T/K)\log U(0,1)$ \Comment{Poisson increments in time}
  \State $k\leftarrow k+1$
  \State $t\leftarrow t+\Delta t$
  \State $t_k\leftarrow t$
  \EndWhile
  \State $\kappa\leftarrow k-1$ \Comment total number of kicks
  \State $t_\mathrm{old}\leftarrow0$
  \For{$k=1,\ldots,\kappa$}
  \State $t_\mathrm{new}\leftarrow t_k$
  \State $(\Delta q)_\mathrm{lg},(\Delta p)_\mathrm{lg}\leftarrow N(0,D_\mathrm{lg}T/K)$ 
\For{$n=1,\ldots,N$}
 \State $(\Delta q)_{n,\mathrm{sm}},(\Delta p)_{n,\mathrm{sm}}\leftarrow N[0,D_\mathrm{sm}(t_\mathrm{new}-t_\mathrm{old})]$ 
   \State $\theta_n\leftarrow
  \theta_n+\Omega(J_n)(t_\mathrm{new}-t_\mathrm{old})$ 
  \State $q_n\leftarrow(2J_n)^{1/2}\sin\theta_n$,
  $p_n\leftarrow(2J_n)^{1/2}\cos\theta_n$
  \State $q_n\leftarrow q_n + (\Delta q)_{n,\mathrm{sm}}+(\Delta
  q)_\mathrm{lg}$ 
\State $p_n\leftarrow p_n + (\Delta p)_{n,\mathrm{sm}}+(\Delta
  p)_\mathrm{lg}$ 
  \State $J_n\leftarrow \half(q_n^2+p_n^2)$,
  $\theta_n\leftarrow\mbox{arctan2}(q_n,p_n)$ 
  \EndFor
  \State $t_\mathrm{old}\leftarrow t_\mathrm{new}$ 
  \EndFor
  \For{$n=1,\ldots,N$} \Comment{add small-scale kicks between $t_\kappa$ and $T$}
 \State $(\Delta q)_{n,\mathrm{sm}},(\Delta p)_{n,\mathrm{sm}}\leftarrow N[0,D_\mathrm{sm}(T-t_\mathrm{old})]$ 
  \State $\theta_n\leftarrow
  \theta_n+\Omega(J_n)(T-t_\mathrm{old})$ 
  \State $q_n\leftarrow(2J_n)^{1/2}\sin\theta_n$,
  $p_n\leftarrow(2J_n)^{1/2}\cos\theta_n$ 
  \State $q_n\leftarrow q_n + (\Delta q)_{n,\mathrm{sm}}$, $p_n\leftarrow p_n + (\Delta p)_{n,\mathrm{sm}}$ 
  \State $J_n\leftarrow \half(q_n^2+p_n^2)$,
  $\theta_n\leftarrow\mbox{arctan2}(q_n,p_n)$ 
  \State Convert $\theta_n$, $J_n$ to position and velocity $z_n$,
  $v_n$
\State \textbf{Return} $z_n,v_n$
   \EndFor
\end{algorithmic}

\begin{figure*}[ht]
\centering
\subfloat[$T-t_1=25$]{
\includegraphics[trim={0.9cm 0.3cm 0.9cm 0.7cm},width=0.47\textwidth]{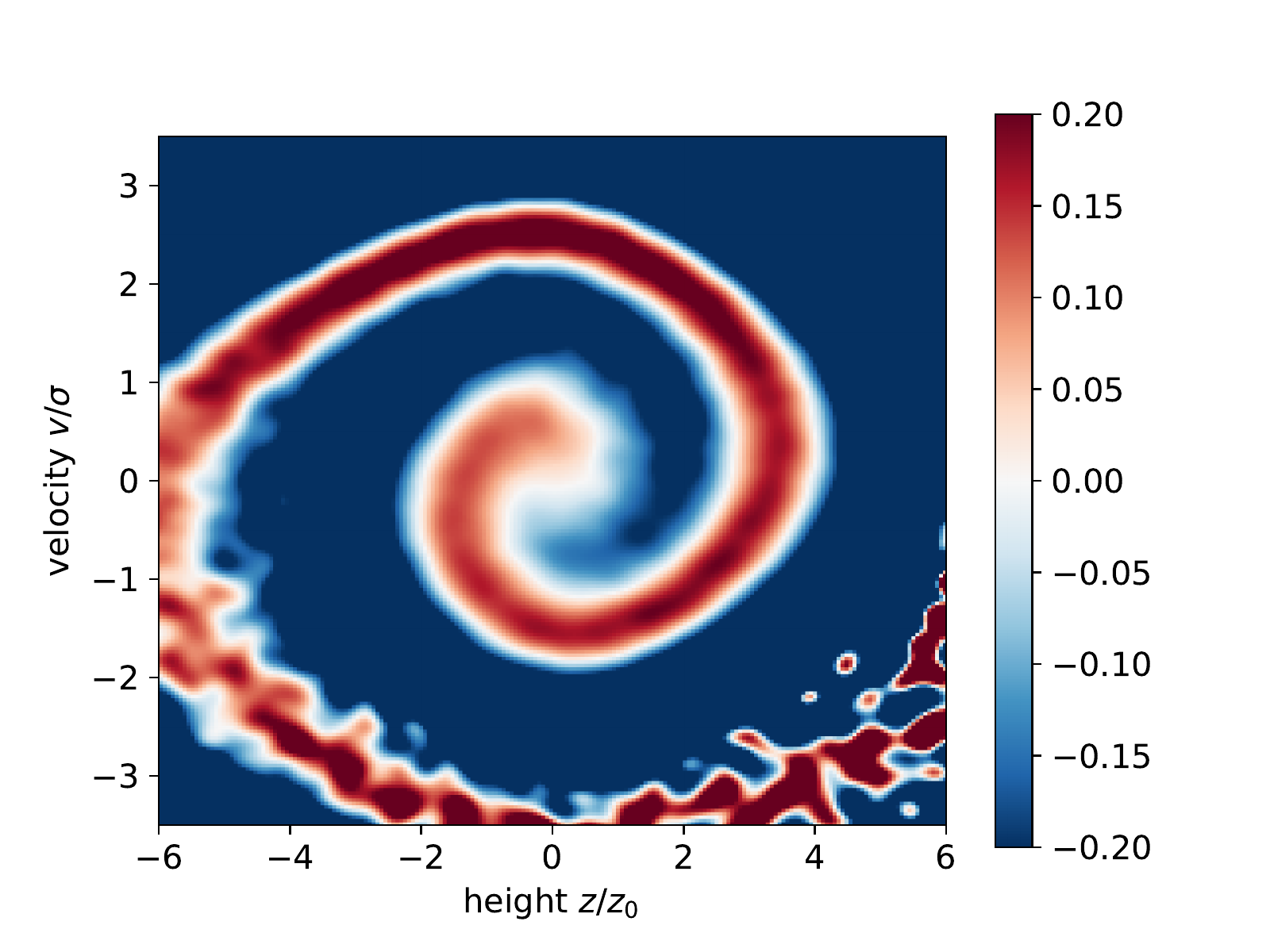}
}\hspace{2pt}
\subfloat[$T-t_1=50$]{
\includegraphics[trim={0.9cm 0.3cm 0.9cm 0.7cm},width=0.47\textwidth]{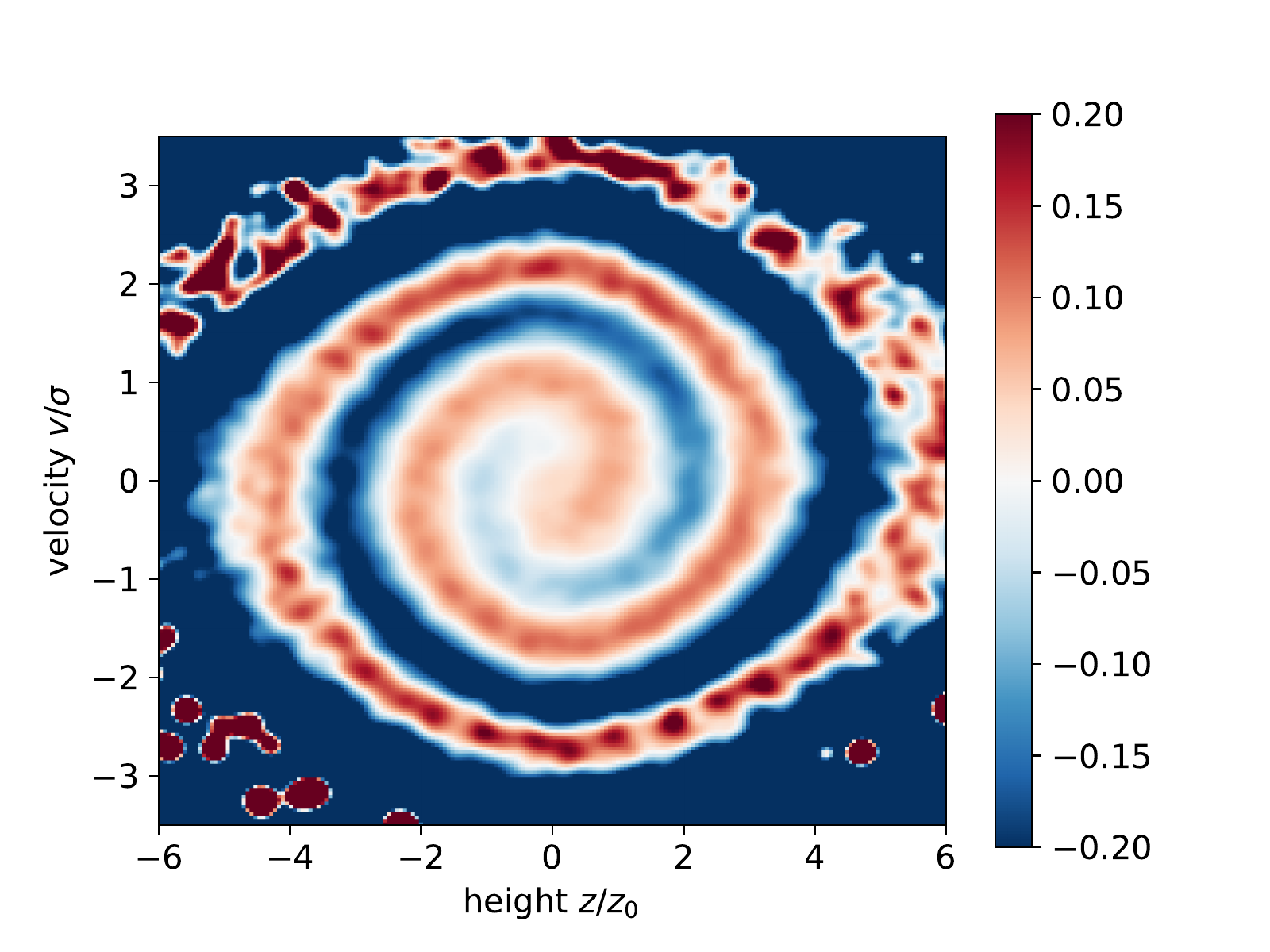}
}
\\
\subfloat[$T-t_1=75$]{
\includegraphics[trim={0.9cm 0.3cm 0.9cm 0.7cm},width=0.47\textwidth]{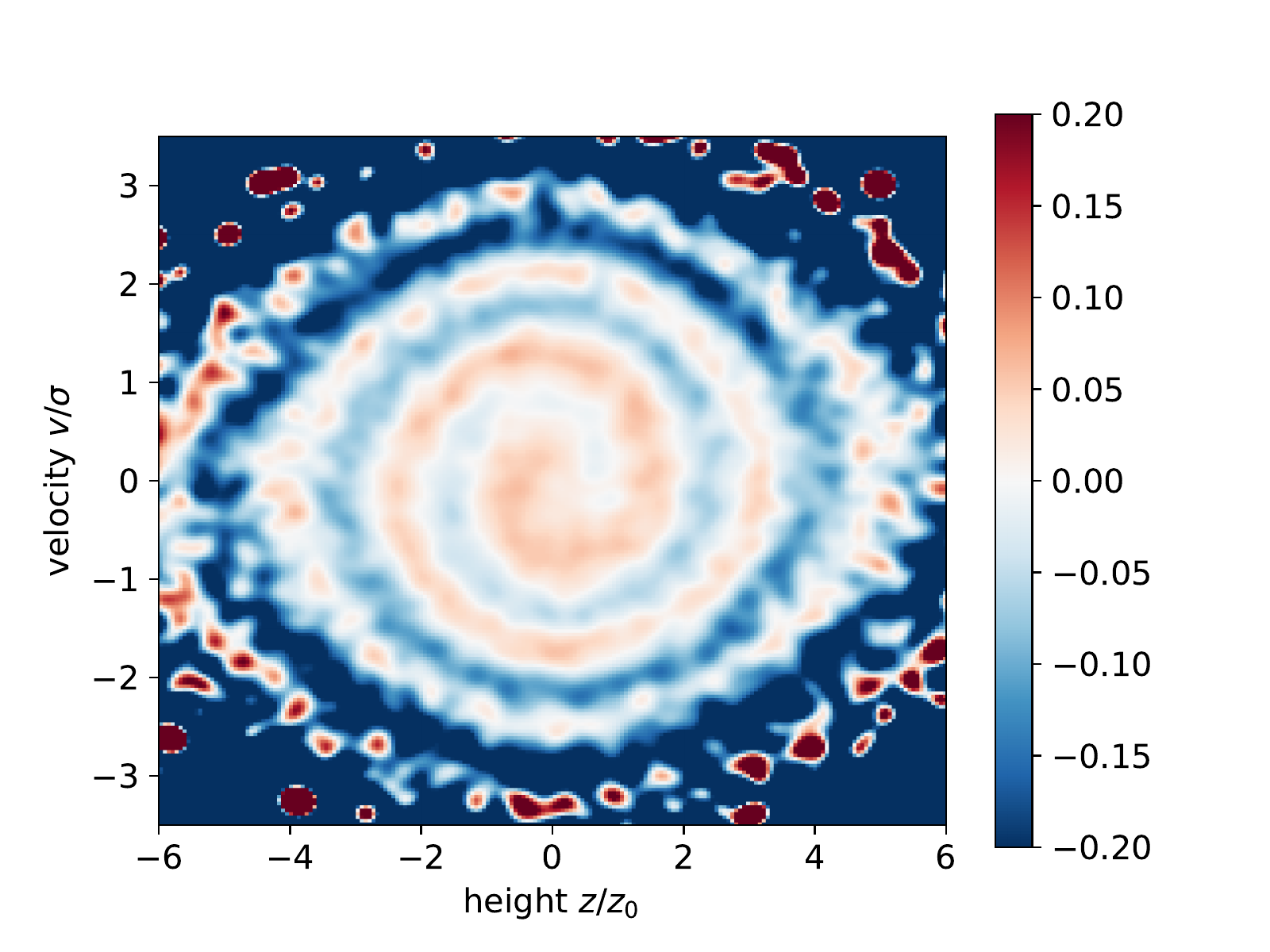}
}
\hspace{2pt}
\subfloat[$T-t_1=100$]{
\includegraphics[trim={0.9cm 0.3cm 0.9cm 0.7cm},width=0.47\textwidth]{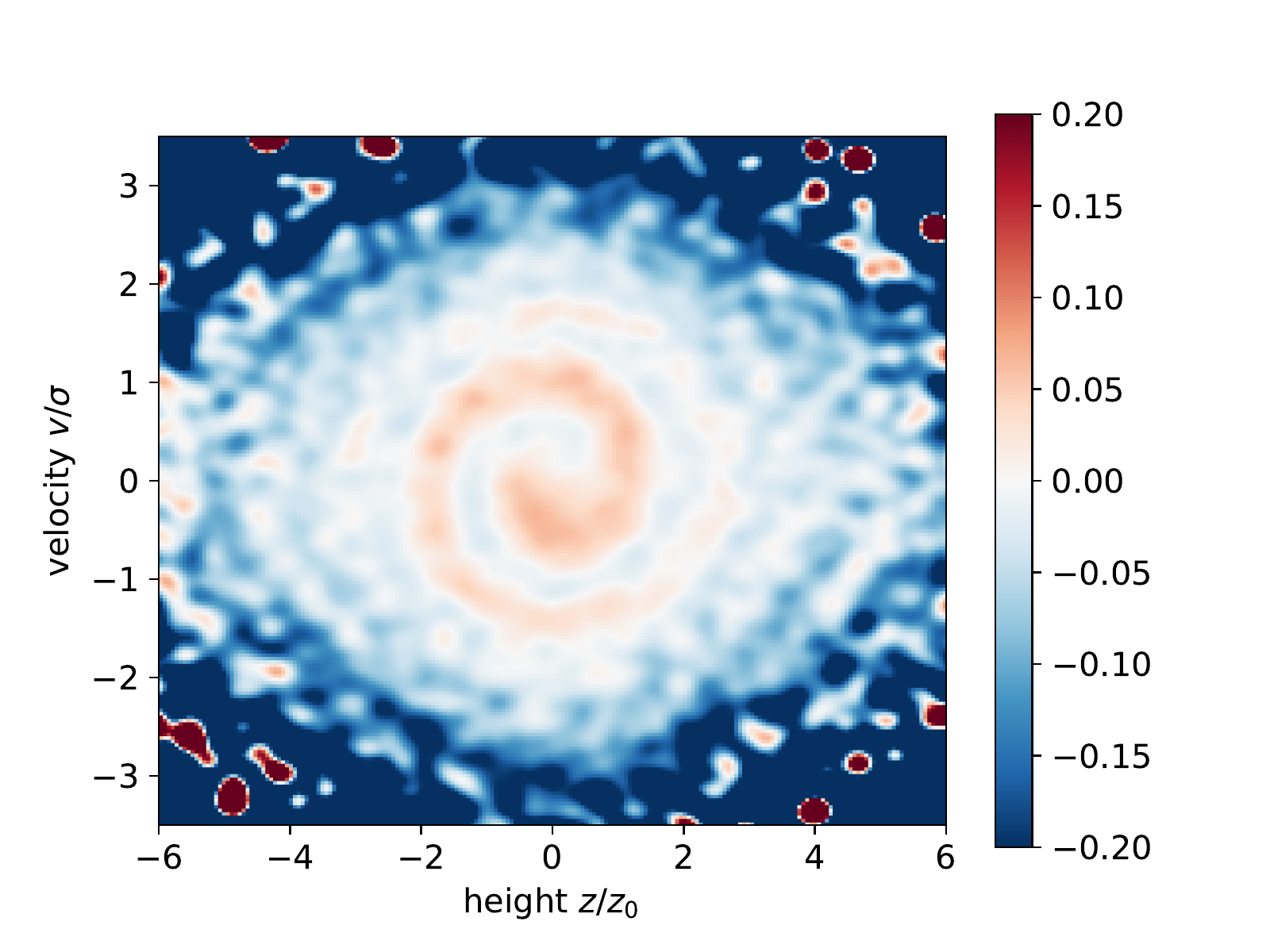}
}
\caption{Simulations of the evolution of structure in the vertical phase space of the solar neighbourhood, arising from a single large-scale kick with an age $T-t_1$. The plots show $\Delta\rho/\overline\rho=(\rho-\overline\rho)/\overline\rho$, where $\rho$ is the phase-space density and $\overline\rho$ is the smoothed density. The fractional contribution from large-scale kicks is $g=0.25$, the disc age $T=1000$, and there are $N=10^6$ stars in the sample.  The event ages in the four panels are 15, 50, 75 and 100; assuming $z_0=300\pc$ and $\sigma=30\kms$ these correspond to ages of 244, 489, 733 and $978\Myr$. Each panel uses a separate random-number seed for all random numbers (the amplitude and direction of the large-scale kick and the small-scale kicks). The snail is mostly washed out by small-scale kicks when the event age is $\gtrsim 1\Gyr$.}
\label{fig:three}
\vspace{10pt}
\end{figure*}

\section{Results}

We have conducted Monte Carlo simulations of the model described in the preceding section. We adopt units in which the current disc's asymptotic scale height $z_0$ and vertical velocity dispersion $\sigma$ are unity. If $z_0=300\pc$ and $\sigma=30\kms$ then the time unit is $9.778\Myr$, so we adopt a disc age $T=1000$, corresponding to $9.78\Gyr$.

All of the simulations reported here have $N=10^6$ stars. The statistical properties of the results of the simulations should depend only on $N$, the disc age $T$, and the fraction $g$ of the heating due to large-scale kicks. The results should be independent of the rate of large-scale kicks, specified by $K$, so long as $K\gg1$. 

\begin{figure*}[ht]
\centering
\subfloat[$g=0$]{
\includegraphics[trim={0.9cm 0.3cm 0.9cm 0.7cm},width=0.47\textwidth]{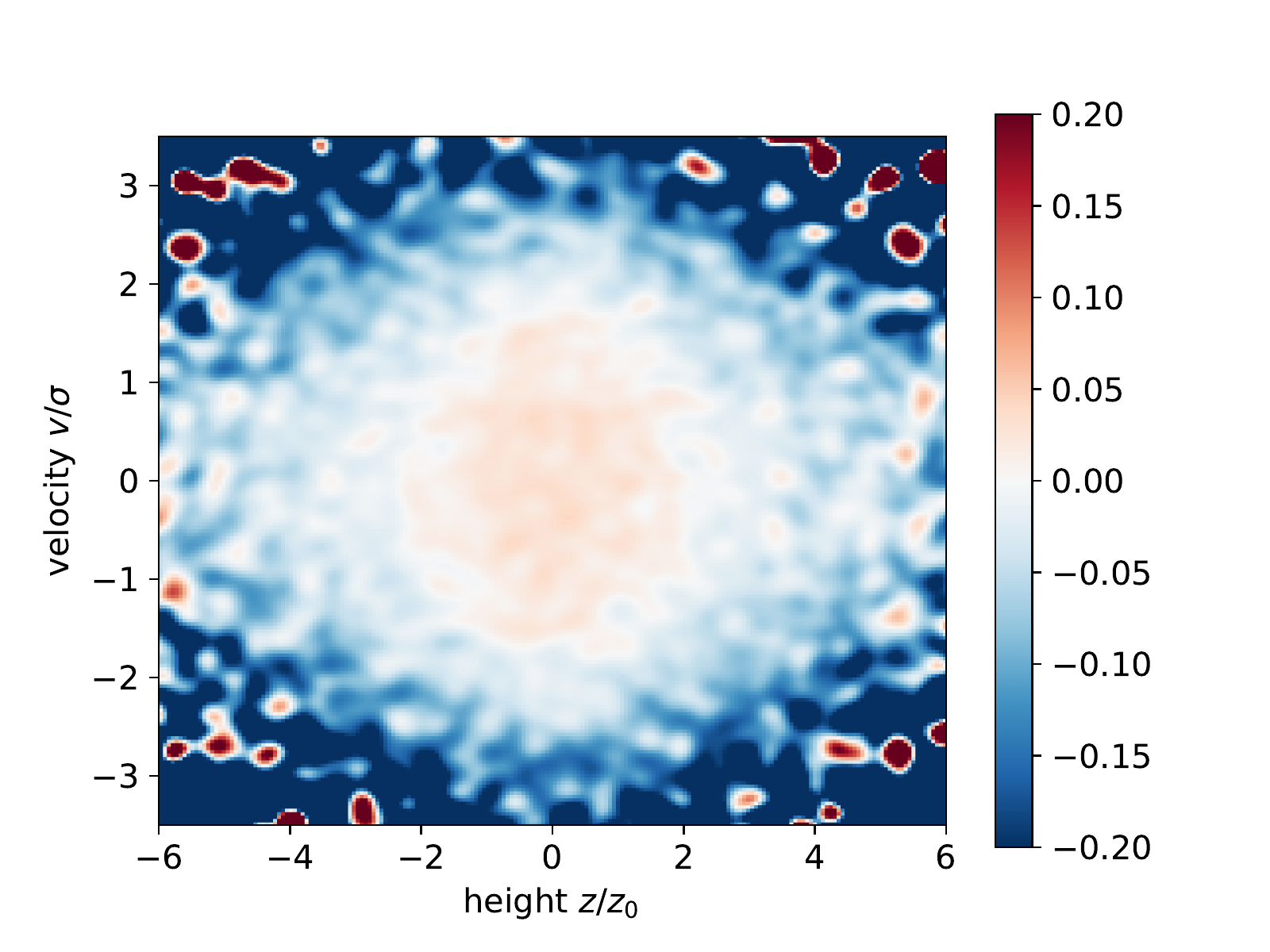}
}\hspace{2pt}
\subfloat[$g=0.25$]{
\includegraphics[trim={0.9cm 0.3cm 0.9cm 0.7cm},width=0.47\textwidth]{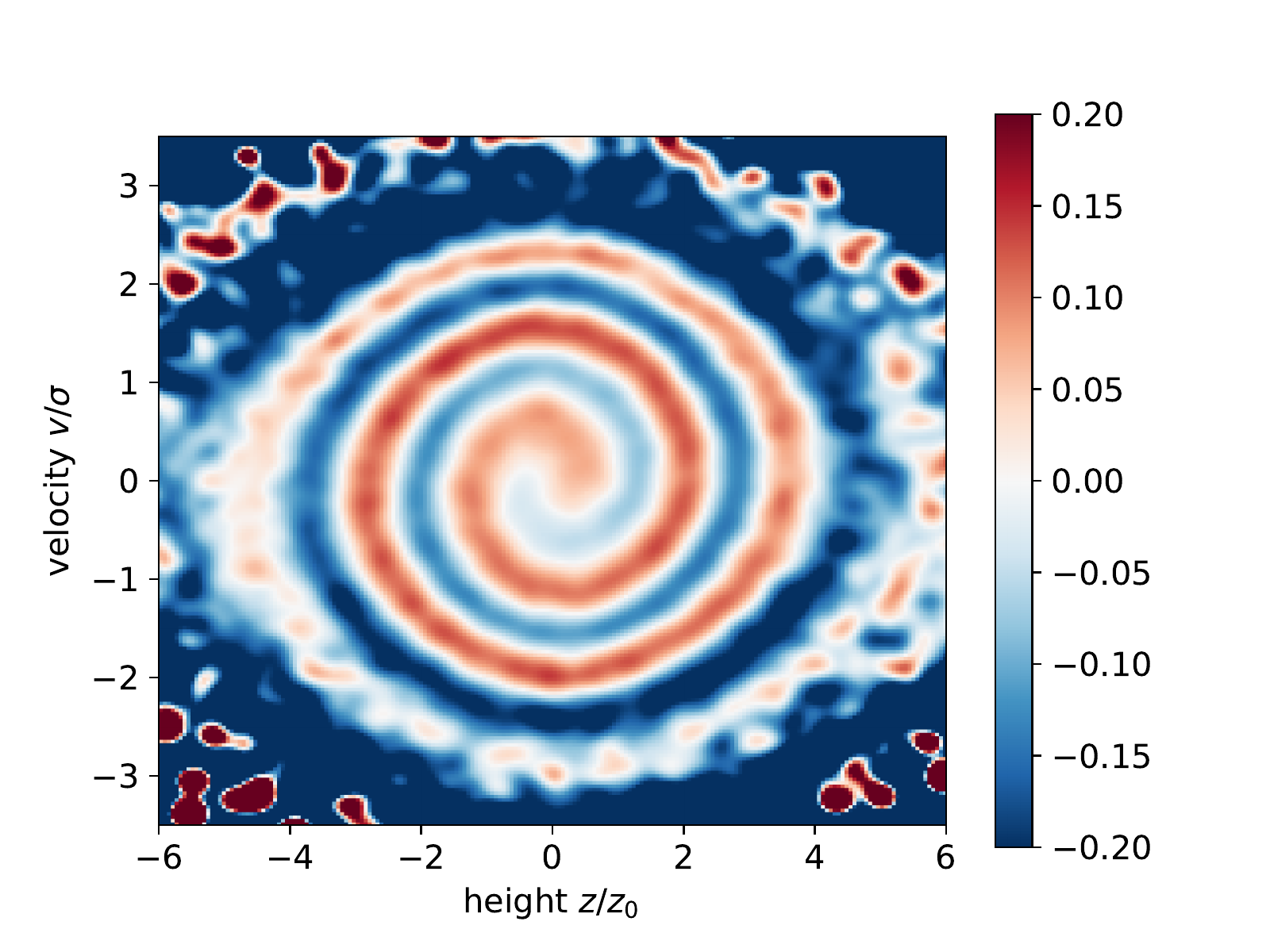}
}
\\
\subfloat[$g=0.5$]{
\includegraphics[trim={0.9cm 0.3cm 0.9cm 0.7cm},width=0.47\textwidth]{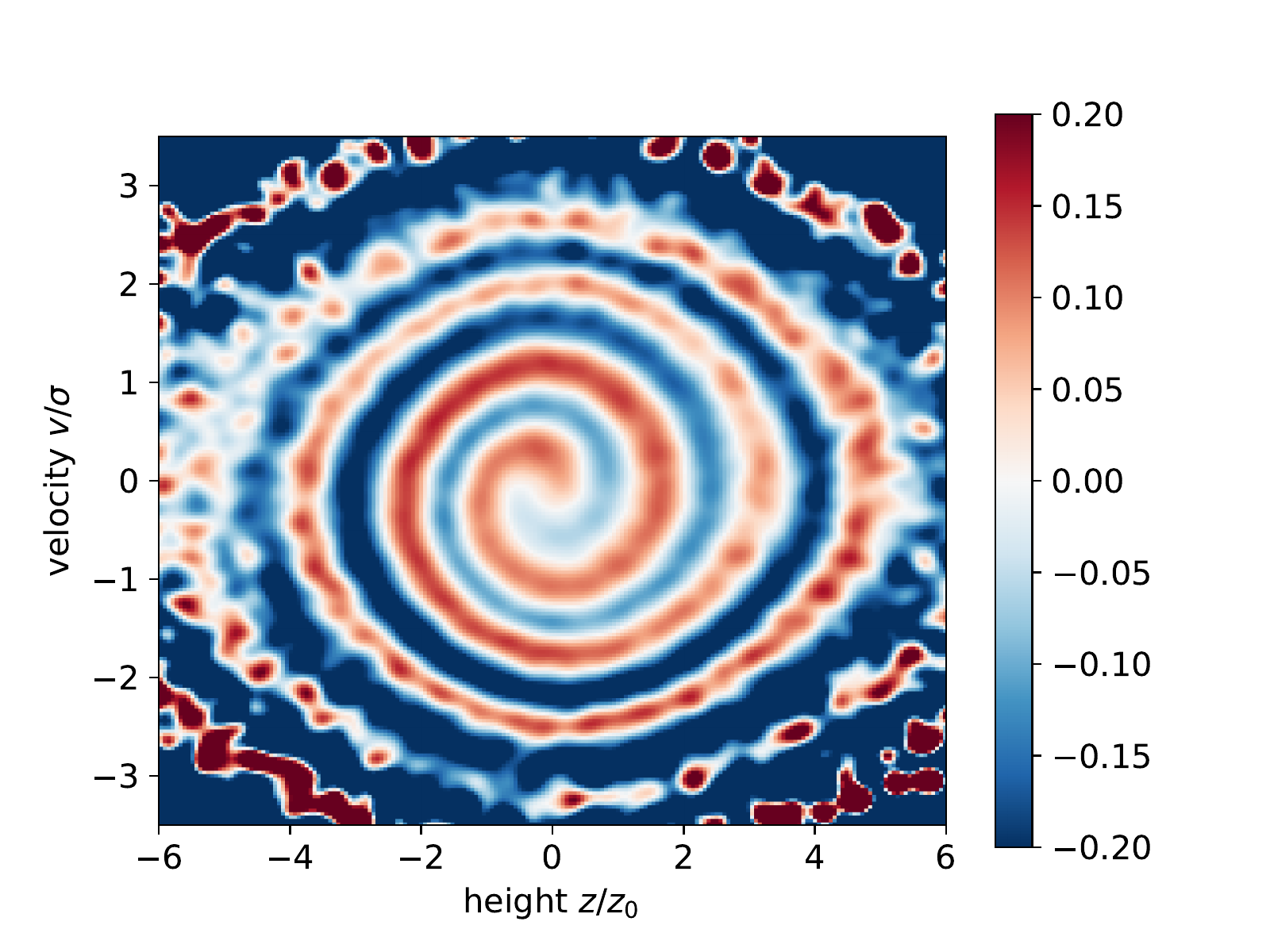}
}
\hspace{2pt}
\subfloat[$g=0.75$]{
\includegraphics[trim={0.9cm 0.3cm 0.9cm 0.7cm},width=0.47\textwidth]{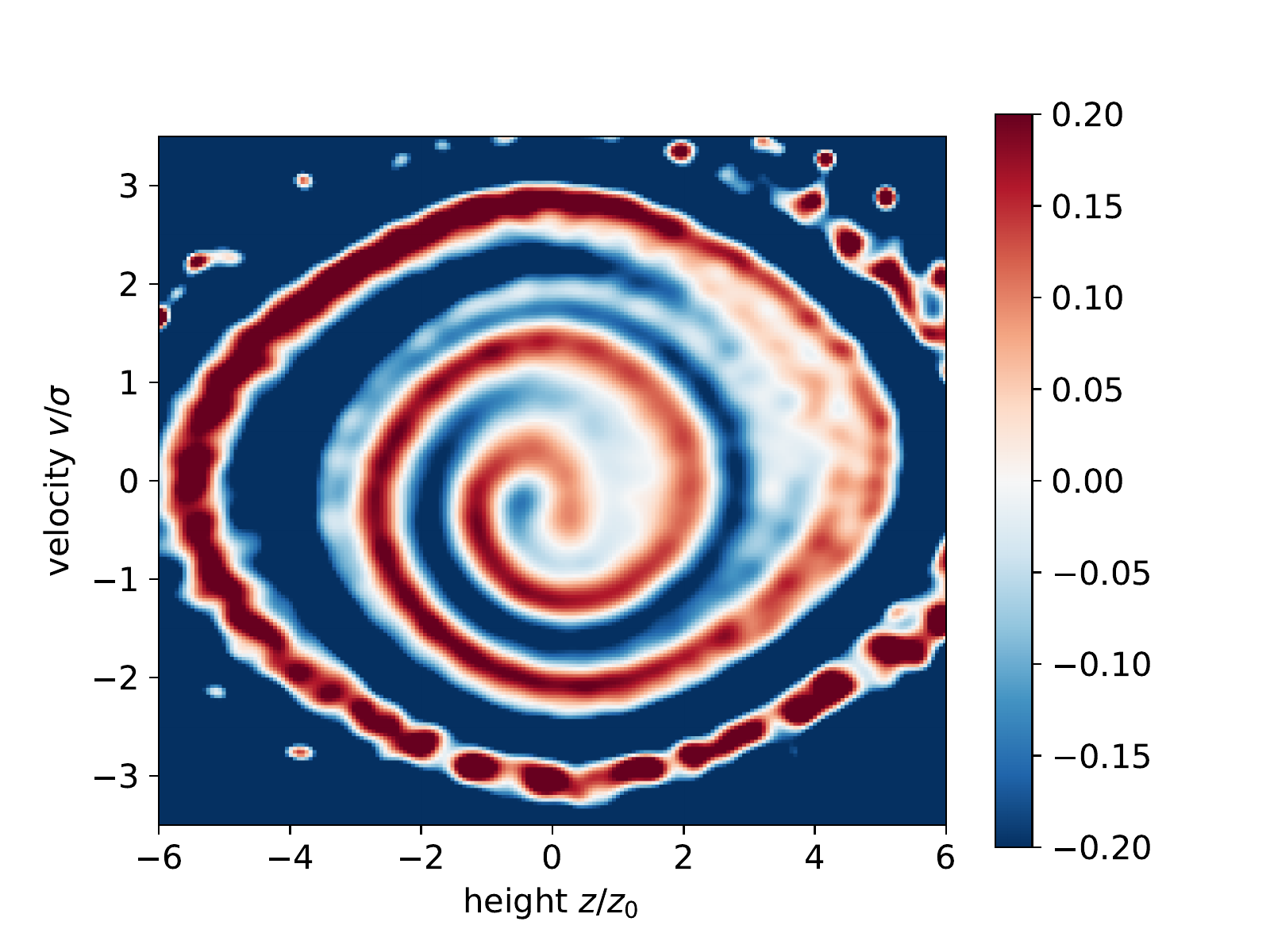}
}
\caption{Simulations of the evolution of structure in the vertical phase space of the solar neighbourhood. The plots show $\Delta\rho/\overline\rho=(\rho-\overline\rho)/\overline\rho$, where $\rho$ is the phase-space density and $\overline\rho$ is the smoothed density. The disc age $T=1000$ and there are $N=10^6$ stars in the sample. The parameter $g$ is the fraction of the squared velocity dispersion that is due to large-scale kicks that affect all the stars, as opposed to small-scale kicks that affect only individual stars. Each panel uses a separate random-number seed for all random numbers (the times of the large-scale kicks and the amplitudes and directions of the large-scale and small-scale kicks).}
\label{fig:one}
\vspace{10pt}
\end{figure*}

\subsection{Destruction of the snail from a single event}

\label{sec:wash}

As a preliminary example, we illustrate how phase-space diffusion from small-scale kicks can destroy the snail produced by a single large-scale kick. To do this, we set the number of large-scale kicks to be $K=1$, choose a fixed time $t_1$ for the large-scale kick instead of choosing this time at random, and study the appearance of the snail as a function of $t_1$.  

The plots in Fig.\ \ref{fig:three} show the vertical phase space -- velocity $v$ versus height $z$, in units of the dispersion $\sigma$ and asymptotic scale height $z_0$. The four panels show kicks with ages $T-t_1=25,50,75$ and 100 (recall that with our fiducial parameters an age of 100 corresponds to $0.98\Gyr$). The plots show $\Delta\rho/\overline\rho=(\rho-\overline\rho)/\overline\rho$, where $\rho$ is the phase-space density and $\overline\rho$ is a smoothed density. Older and older snails are more and more tightly wound but snails older than $\sim 1\Gyr$ are mostly washed out by the cumulative effects of small-scale kicks, as described in the Introduction. 

A striking feature of Fig.\ \ref{fig:three} is the rapid damping of the snail after an age $T-t_1=50$. This behavior arises because the amplitude of the snail decays as $F(\tau)=\exp(-\tau^3/t_\mathrm{max}^3)$ where $\tau$ is the age of the snail (eq.\ \ref{eq:dampdef}). At the mean action $\langle J\rangle=1.7974$, equation  (\ref{eq:tmaxmean}) implies that $t_\mathrm{max}\simeq 80$, so in the last two panels we expect the amplitude of the snail to be damped from its original value by factors $F(75)=0.44$ and $F(100)=0.14$. 

Note that at $\tau=100$ the snail is strongest near the centre, where the actions are small, the shear $\Gamma$ is nearly zero, and $t_\mathrm{max}\propto \Gamma^{-2/3}$ is large. This behavior may not be present in more realistic models of the vertical potential, where the shear is larger at small actions (see Fig.\ \ref{fig:shear}).

\subsection{Snails from a Gaussian process }

\begin{figure*}[ht]
\centering
\subfloat[$g=0$]{
\includegraphics[trim={0.9cm 0.3cm 0.9cm 0.7cm},width=0.47\textwidth]{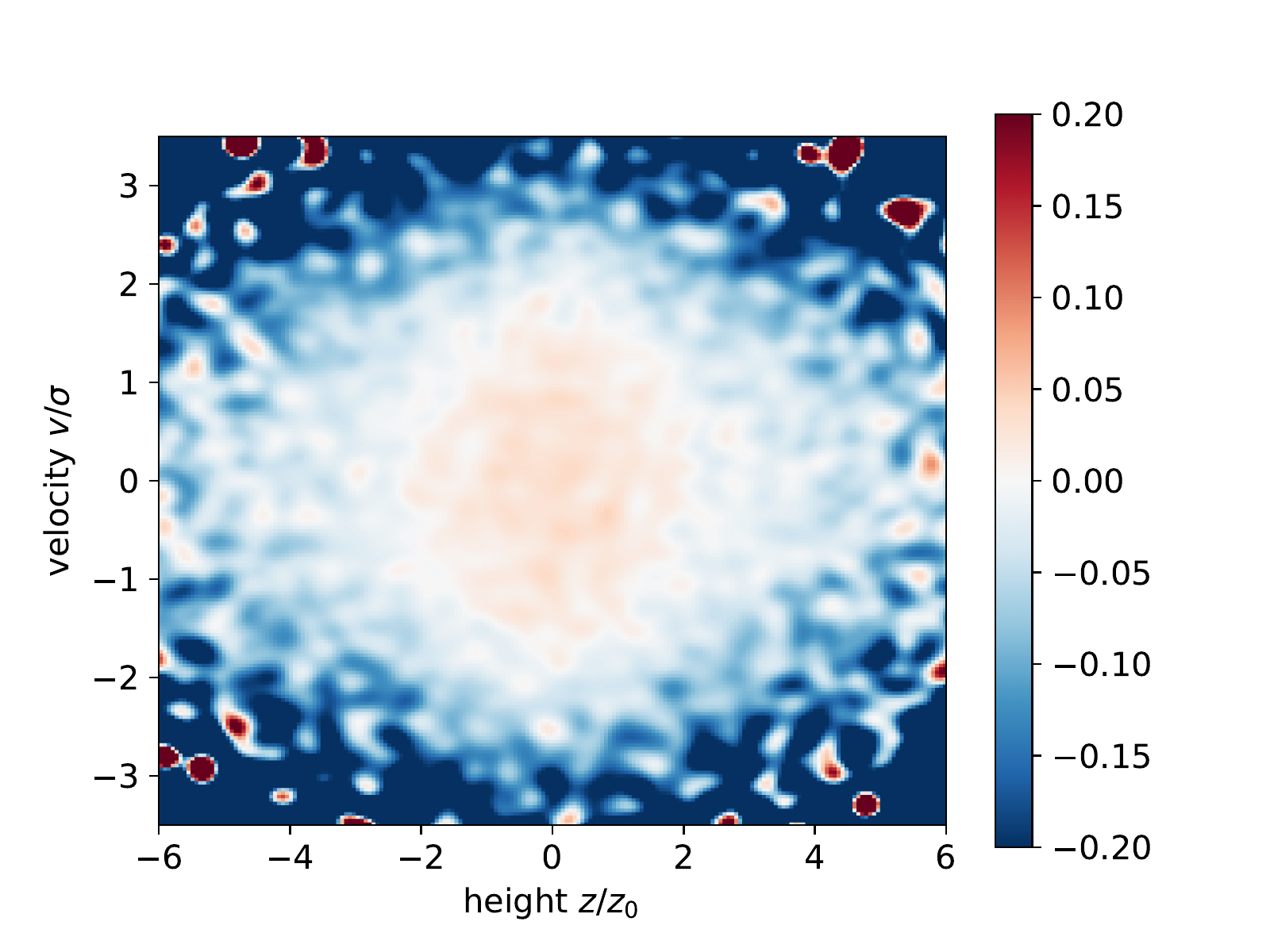}
}\hspace{2pt}
\subfloat[$g=0.25$]{
\includegraphics[trim={0.9cm 0.3cm 0.9cm 0.7cm},width=0.47\textwidth]{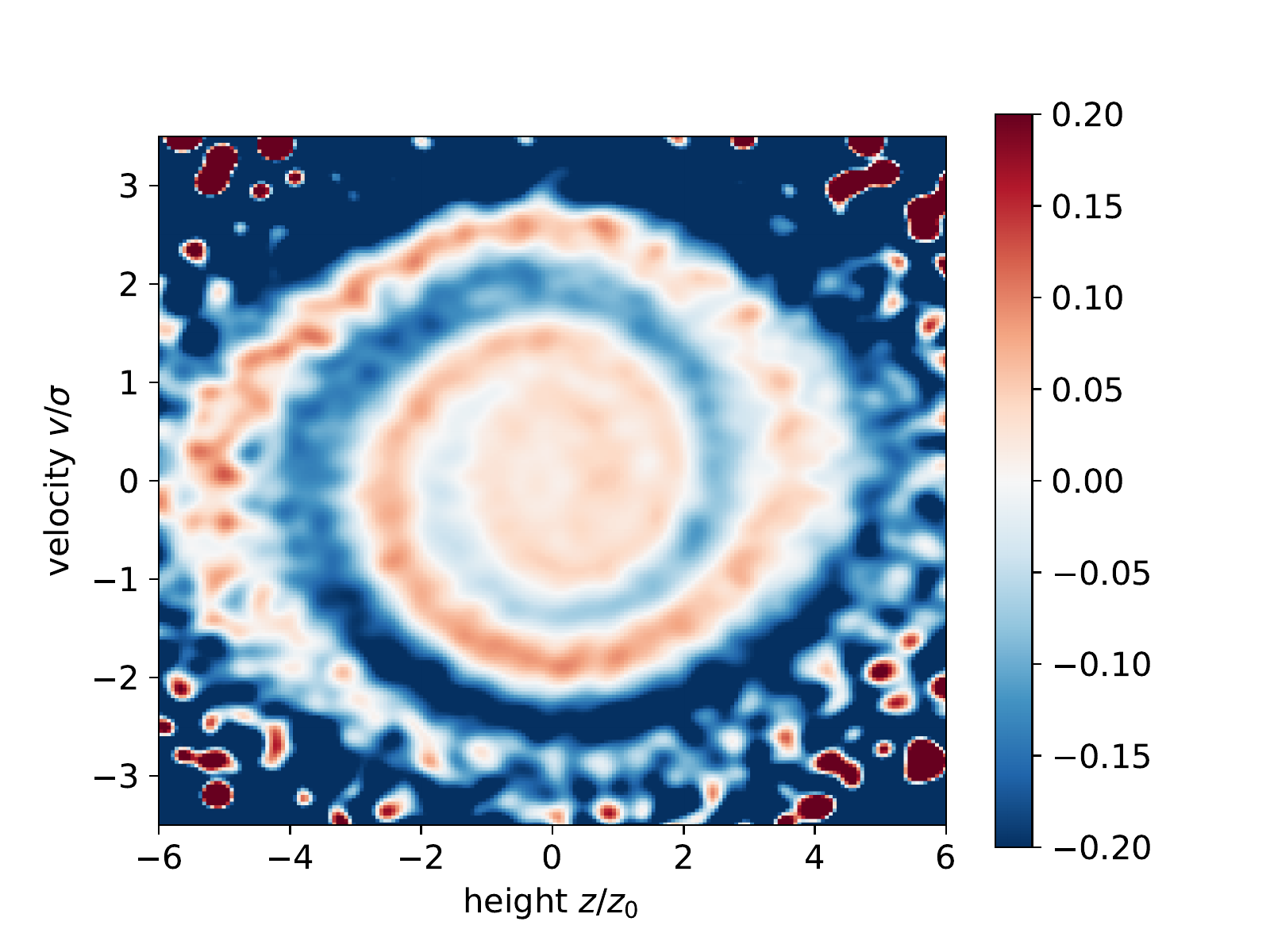}
}
\\
\subfloat[$g=0.5$]{
\includegraphics[trim={0.9cm 0.3cm 0.9cm 0.7cm},width=0.47\textwidth]{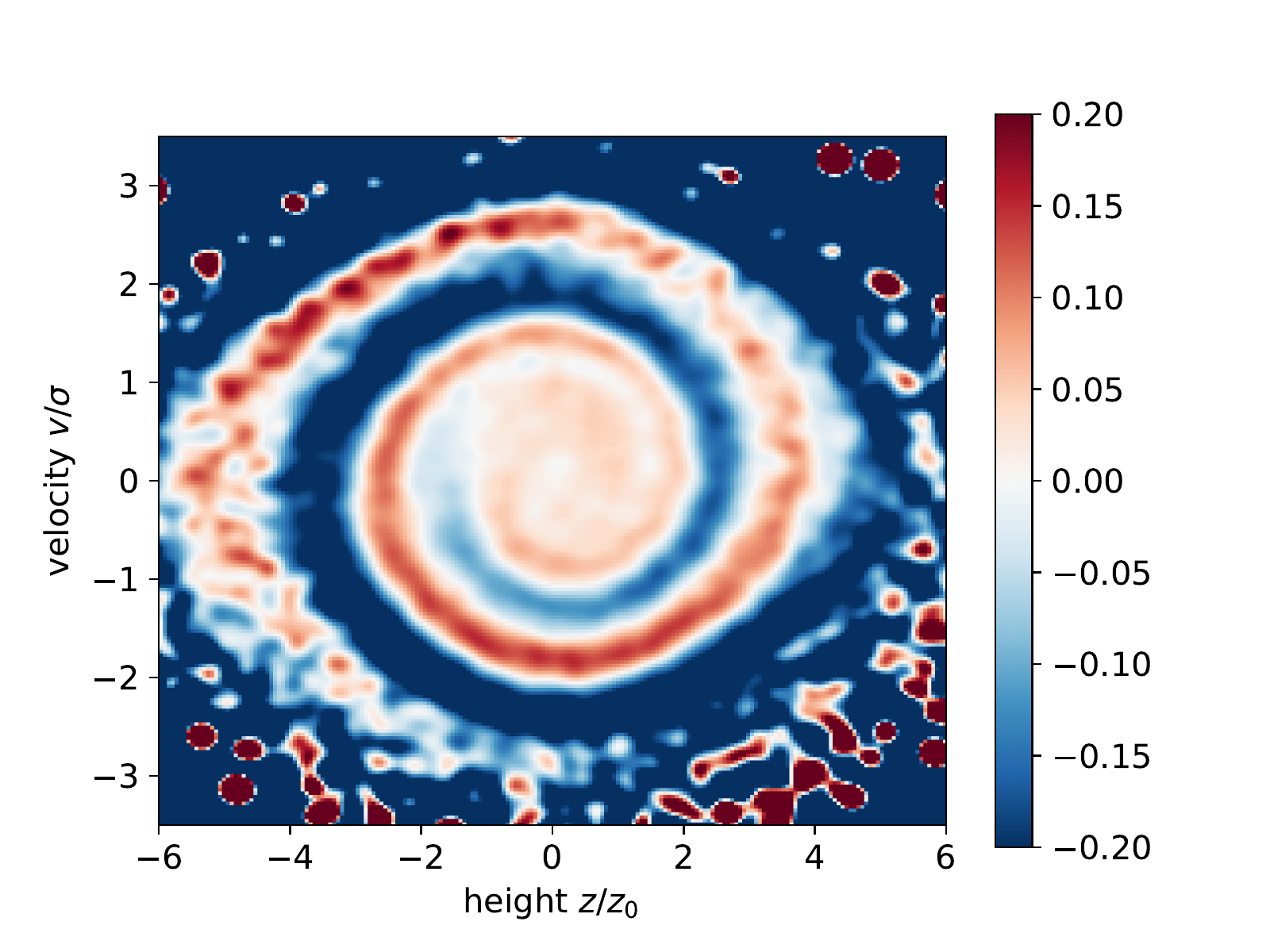}
}
\hspace{2pt}
\subfloat[$g=0.75$]{
\includegraphics[trim={0.9cm 0.3cm 0.9cm 0.7cm},width=0.47\textwidth]{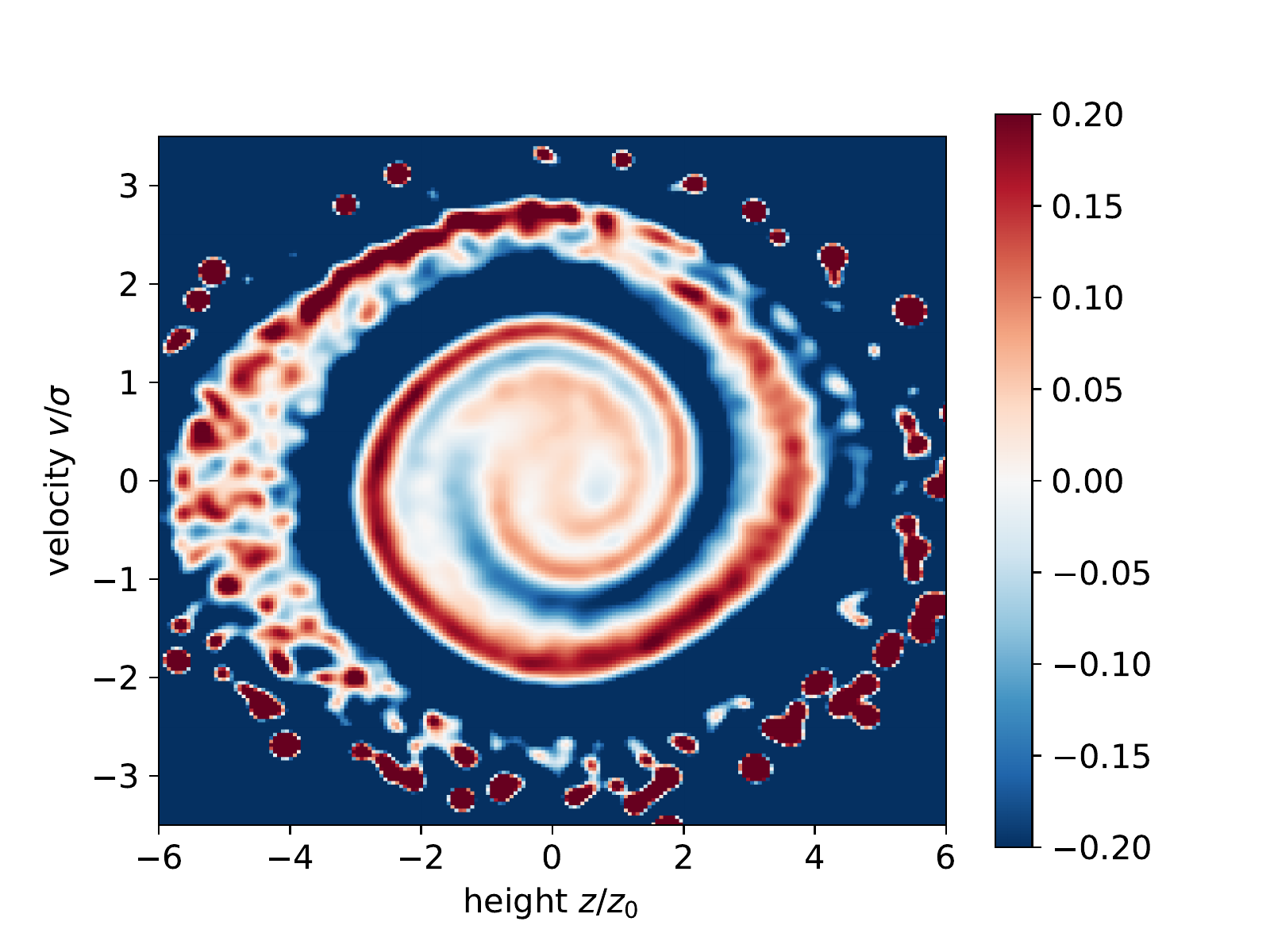}
}
\caption{As in Fig.\ \ref{fig:one}, but with the same random-number seed for each panel.}
\label{fig:neige}
\vspace{10pt}
\end{figure*}

The four panels in Fig.\ \ref{fig:one} show the results of simulations with an average of $K=100$ large-scale kicks and $g=0$ (no large-scale kicks), $g=0.25$, $g=0.5$ and $g=0.75$. All of the simulations with $g>0$ exhibit prominent spiral structure reminiscent of the Gaia snail, even though no one large-scale kick dominates the perturbations. To confirm that this is not an unusual accident, Fig.\ \ref{fig:neige} shows similar simulations with a different random-number seed that is the same for all four panels (thus the times of the large-scale kicks and the direction of the large-scale and small-scale kicks are the same in each panel).

The amplitude of the snail in the simulations is somewhat larger than the amplitude in the data (Fig.\ \ref{fig:snail}), probably because our model has only one degree of freedom and thus does not account for the weakening of the snail as it shears out in azimuth. 

\begin{figure*}[ht]
\centering
\subfloat[$g=0$]{
\includegraphics[trim={0.9cm 0.3cm 0.9cm 0.7cm},width=0.47\textwidth]{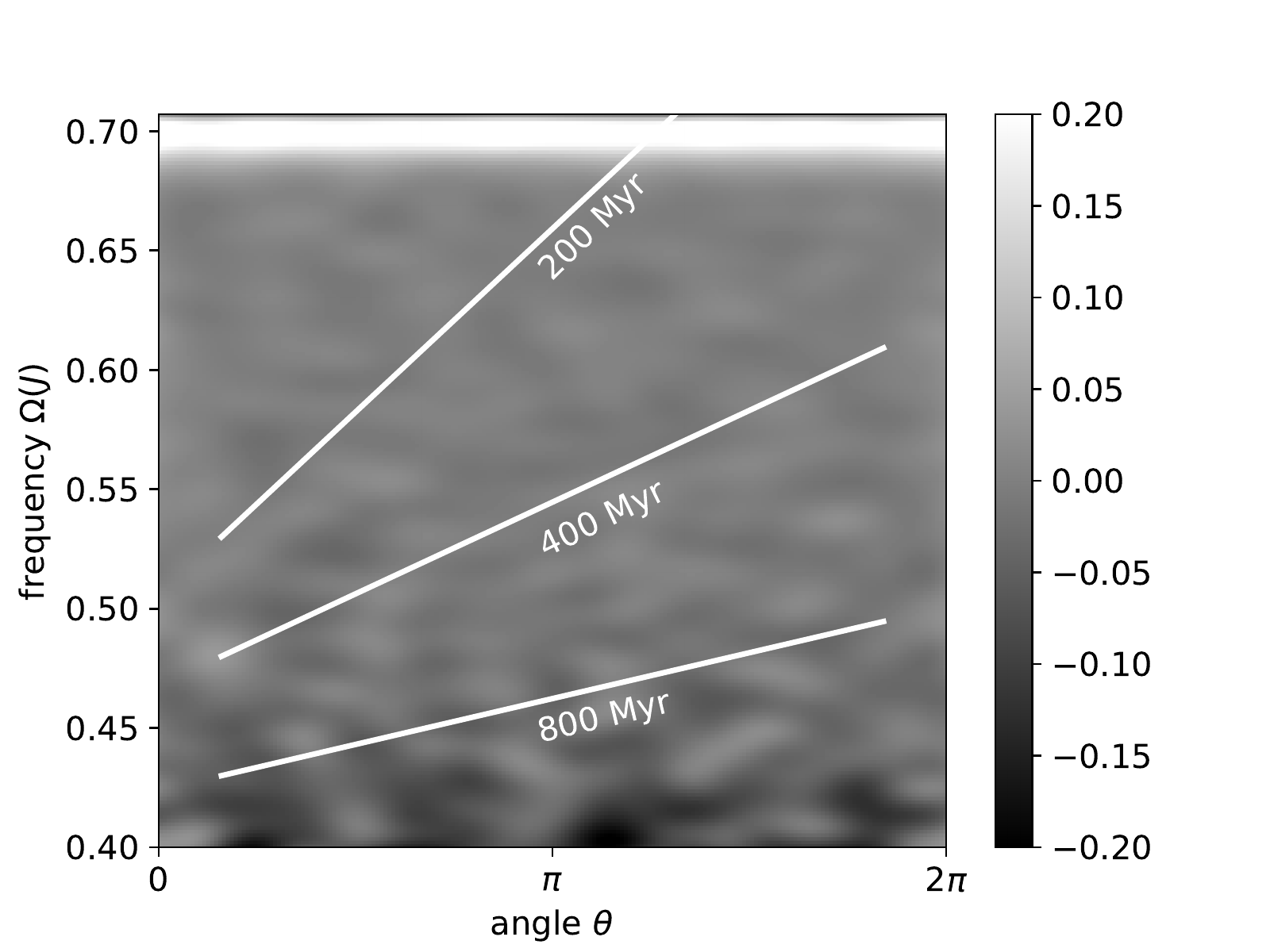}
}\hspace{2pt}
\subfloat[$g=0.25$]{
\includegraphics[trim={0.9cm 0.3cm 0.9cm 0.7cm},width=0.47\textwidth]{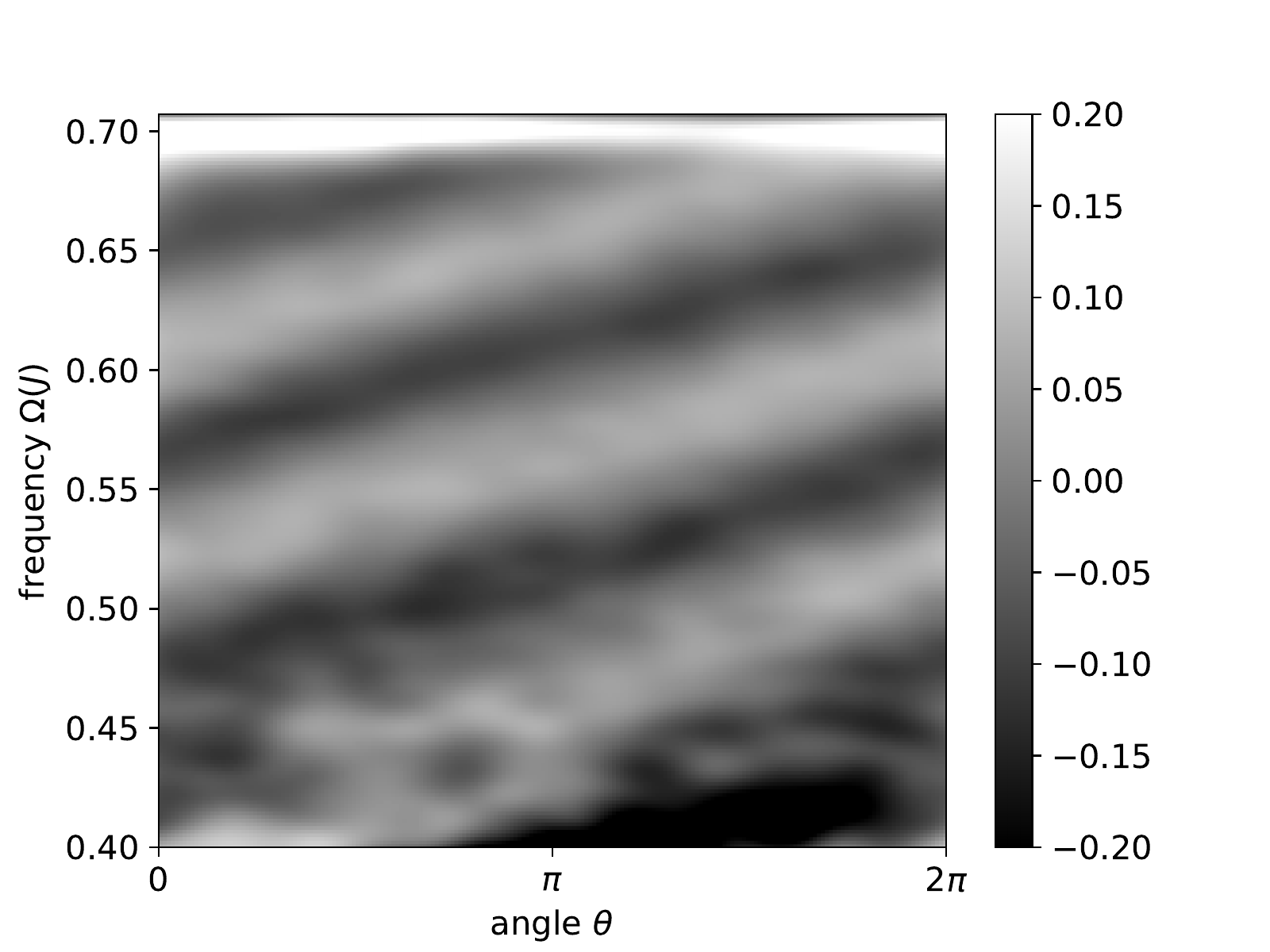}
}
\\
\subfloat[$g=0.5$]{
\includegraphics[trim={0.9cm 0.3cm 0.9cm 0.7cm},width=0.47\textwidth]{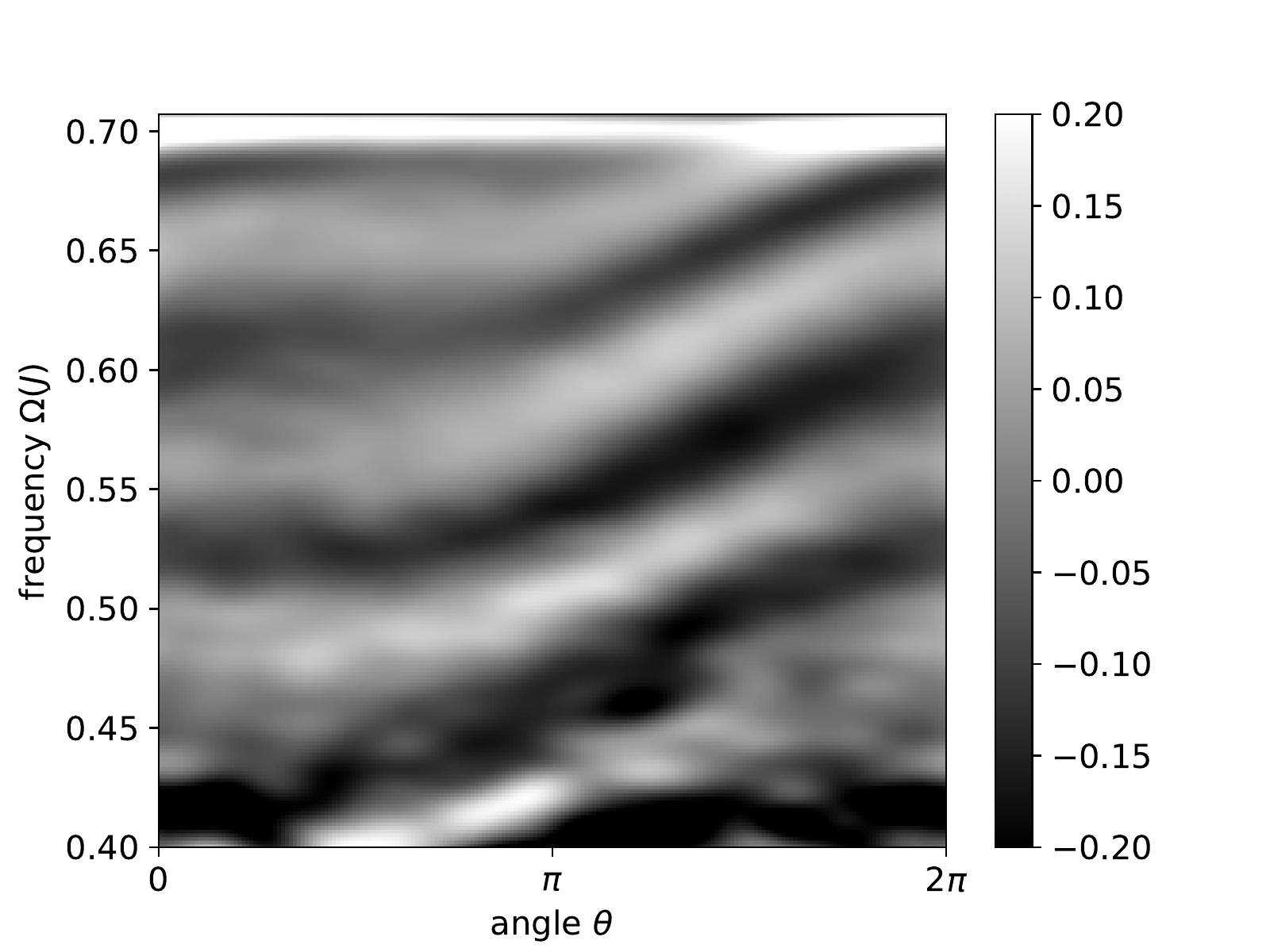}
}
\hspace{2pt}
\subfloat[$g=0.75$]{
\includegraphics[trim={0.9cm 0.3cm 0.9cm 0.7cm},width=0.47\textwidth]{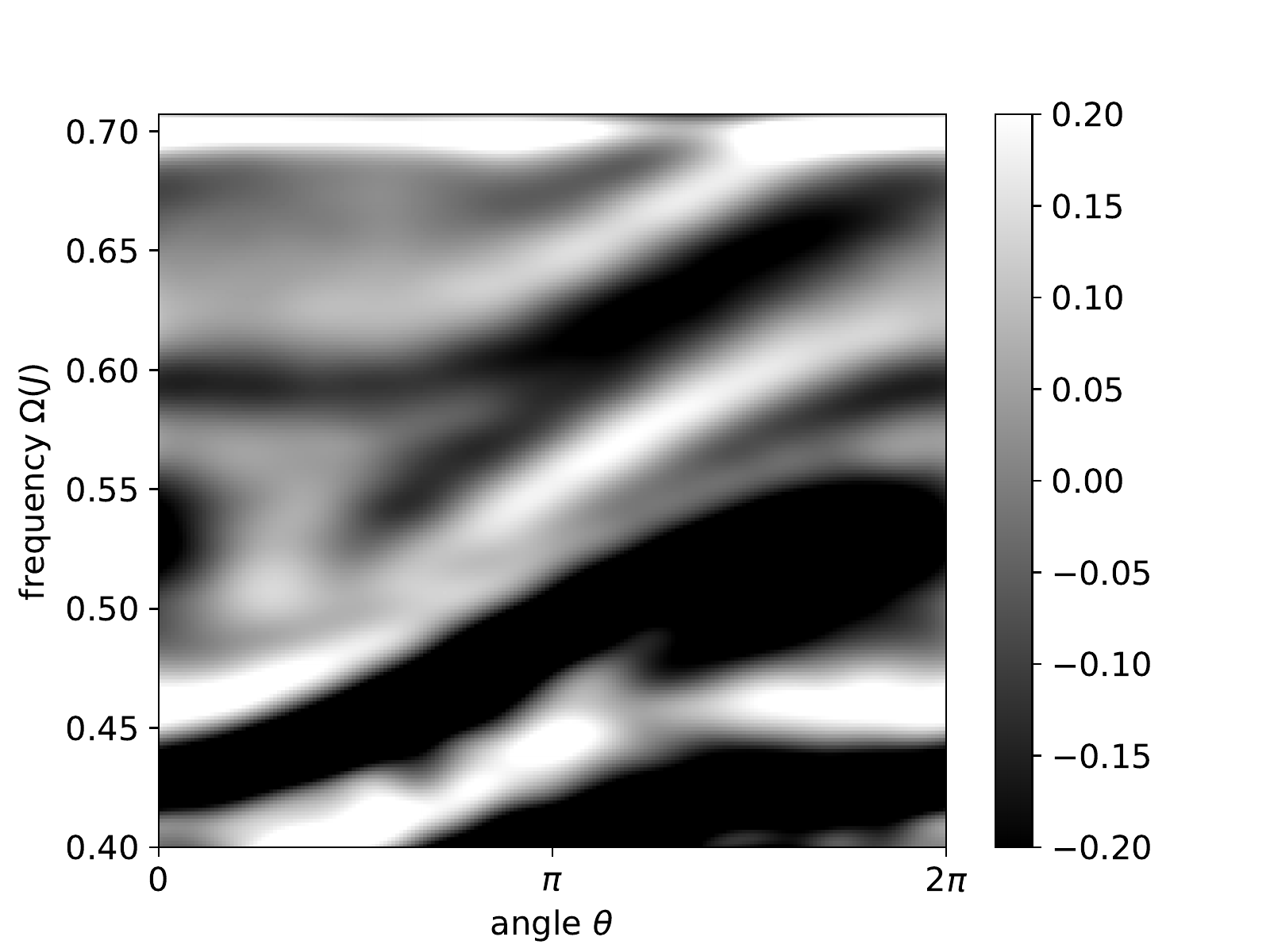}
}
\caption{As in Fig.\ \ref{fig:one}, except the plots show $\Delta\rho/\overline\rho$ in the space of angle $\theta$ and frequency $\Omega(J)$. Perturbations that are localized in angle and time would excite responses that appear as straight lines in these plots, with a slope equal to the inverse of the age of the perturbation event. The slopes for several event ages are shown in the top left panel, assuming $z_0=300\pc$ and $\sigma=30\kms$. The panels appear to show evidence that the structures are mainly due to a single event with an age $\sim0.5\Gyr$ but in fact they are created by Gaussian noise. }
\label{fig:two}
\vspace{10pt}
\end{figure*}

The plots in Fig.\ \ref{fig:two} show the fractional density contrast in the $\theta$--$\Omega(J)$ plane (the `zebra' diagram). An instantaneous perturbation to the disc at time $t_i$ and angle $\theta_i$ should be visible at the present time $T$ as the straight line $\Omega=(\theta-\theta_i)/(T-t_i)$. Thus the slopes of straight features in these plots are often interpreted as the times at which these features were formed. Some typical event times are shown in the top left panel. All of the panels with $g>0$ exhibit features that look more-or-less straight and could be (erroneously) interpreted to mean that a single event was mostly responsible for the snail. In fact the sloped features are a property of Gaussian noise that is band-limited at the frequency $1/t_\mathrm{max}$. 

This argument can be made (somewhat) more quantitative. At a fixed angle $\theta=\theta'$ the correlation function (\ref{eq:xi}) simplifies to 
\begin{align}
    \langle \Delta f(J,\theta)&\Delta f(J',\theta)\rangle
    =\half r\langle \Delta f(J)\Delta f(J')\rangle\nonumber \\ &\times \frac{\sin(\Omega-\Omega')t_\mathrm{max}}{\Omega-\Omega'}.
    \label{eq:xi2}
\end{align}
When $\Omega t_\mathrm{max}\gg1$, which is the case in the solar neighbourhood, we can ignore the slow variation of the factor $\langle \Delta f(J)\Delta f(J')\rangle$ and write
\begin{equation}
    \langle \Delta f(J,\theta)\Delta f(J',\theta)\rangle
    =\mbox{const}\times\frac{\sin[(\Omega-\Omega')t_\mathrm{max}]}{\Omega-\Omega'}.
    \label{eq:xi3}
\end{equation}
This equation describes the correlation function of a Gaussian process on a line parallel to the $\Omega$-axis that is band-limited at $t_\mathrm{max}$. The frequency of maxima (average number of maxima per unit change in $\Omega$) in such a process is $(\frac{3}{5})^{1/2}(2\pi)^{-1}t_\mathrm{max}$ \citep{rice45}. In contrast, if the snail is due to a single event of age $\tau$, the density at a given angle varies as $\cos[\Omega(J)\tau+\mbox{const}]$ (cf.\ eq.\ \ref{eq:wrap}) so the frequency of maxima is $(2\pi)^{-1}\tau$. These results suggest that the eye will tend to find linear features in plots like Fig.\ \ref{fig:two}) with slopes corresponding roughly to an age $\tau=(\frac{3}{5})^{1/2}t_\mathrm{max}=0.77\,t_\mathrm{max}$. For the estimate $t_\mathrm{max}\simeq 0.6\Gyr$ from \S\ref{sec:snaildyn} we find $\tau\simeq 0.45\Gyr$, consistent with the appearance of Fig.\  \ref{fig:two}. This calculation involves \emph{no} free parameters, and is consistent with the range of published estimates of the age of the hypothetical encounter that excited the snail (see footnote \ref{foot:one}). 

\section{Discussion and summary}

We have shown that phase-space features similar to the Gaia snail can be generated by a Gaussian process consisting of many weak perturbations to the distribution function distributed randomly in time. The characteristic age of the snail, determined by its degree of winding, is determined by the age at which small-scale perturbations due to giant molecular clouds or other masses erase the snail, rather than the time of an accidental close encounter with a single perturber. 

Our model depends on the assumption that the combined effects of many weak perturbations are stronger than the effects of the one or two strongest perturbations to the disc. Suppose that the perturbations are due to sub-halos and the number density of sub-halos with masses between $m$ and $2m$ is $n(m)\propto m^{1-\alpha}$. Then if the disturbances add incoherently, the contribution to the total perturbation to the distribution function from sub-halos in this mass range is $\propto mn^{1/2}(m)\propto m^\beta$ with $\beta\equiv (3-\alpha)/2$. Simulations of structure formation in $\Lambda$CDM cosmologies suggest $\alpha\simeq 1.9$ \citep{springel08}, in which case $\beta=0.55$ and the effects of the most massive sub-halos would dominate the perturbations, a result that is confirmed by simulations \citep{mj2016,grand2016}. A significant fraction of the sub-halos orbiting close to the solar radius are tidally disrupted, but tides appear to deplete high-mass and low-mass halos at roughly the same fractional rate \citep{wb20} and thus do not modify this argument. If this theoretical model for the distribution of sub-halos is correct, then models of the snail based on an encounter with a single massive sub-halo such as the Sagittarius dwarf may be more appropriate than models based on Gaussian noise. In practice, these are two extremes of a continuum (see for example \citealt{hunt22}), and the most realistic models may involve a combination of the two. 

The simulations in this paper use a simple model that assumes slab symmetry, in which the phase space has only one degree of freedom. A more realistic model would include the rotational and radial motion of the stars in the disc plane as well as the observational selection effects in the Gaia sample, primarily the restriction of the sample to stars close to the Sun. In particular, large-scale perturbations to the disc stars will initially be localized to a portion of the disc and will weaken with time as they spread out due to differential rotation \citep{banik22}. This process also contributes to erasing old snails. 

Our model is intended to describe the properties of the snail in the solar neighbourhood. At larger distances from the Galactic centre the surface density of giant molecular clouds is much smaller and the orbital periods are longer. In these regions the approximation of the excitation of the disc as a Gaussian process is probably not accurate and it may be possible to identify the times and other properties of specific large-scale kicks that excited the snail.

We have modelled the effect of a perturbation on the distribution function as a change in the canonical coordinates and momenta $(\Delta q,\Delta p)$ that is a Gaussian random variable with zero mean, independent of the initial phase-space position. A more realistic model would account for the dependence of $(\Delta q,\Delta p)$ on $(q,p)$ and on (i) the impact parameter, velocity, and orbital inclination of the perturber \citep{banik22}, for large-scale perturbations; (ii) the properties of the population of giant molecular clouds or other sources of small-scale perturbations.

Simulations of a disc of test particles perturbed by multiple passing sub-halos and a population of small-scale perturbers would address many of these issues.

Our results imply that models in which the snail arises from a single event may be misleading. A promising direction for future research would be to model the snail using the machinery of Gaussian processes and correlation functions, both the correlation function of potential perturbations in the Milky Way $\langle\Phi({\bf x},t)\Phi({\bf x}',t')\rangle$, which describes the excitation, and the correlation function of the phase-space distribution function, $\langle f({\bf x},{\bf v})f({\bf x}',{\bf v}')\rangle$, which describes the response. 

\section*{Acknowledgements}

This work has made use of data from the European Space Agency (ESA) mission {\it Gaia} (\url{https://www.cosmos.esa.int/gaia}), processed by the {\it Gaia} Data Processing and Analysis Consortium (DPAC, \url{https://www.cosmos.esa.int/web/gaia/dpac/consortium}). Funding for the DPAC has been provided by national institutions, in particular the institutions participating in the {\it Gaia} Multilateral Agreement. This work was also supported in part by the Natural Sciences and Engineering Research Council of Canada (NSERC), funding reference numbers RGPIN-2020-03885, CITA 490888-16, and RGPIN-2020-04712. NF acknowledges partial support from an Arts \& Sciences Postdoctoral Fellowship at the University of Toronto.

\section*{Data availability}
The observational data underlying this article were accessed from the Gaia archive (https://gea.esac.esa.int/archive/).
The data underlying this article will be shared on reasonable request to the corresponding author.

\appendix

\section{}

This appendix presents an analytic derivation of the late-time behavior of the snail in the presence of small-scale kicks.

In action-angle variables $(J,\theta)$ the evolution of the distribution function $f(J,\theta,t)$ is described by the Boltzmann equation
\begin{equation}
    \frac{\p f}{\p t} + \Omega(J)\frac{\p f}{\p\theta}=C[f],
\end{equation}
where $\Omega(J)$ is given by equation (\ref{eq:omega}) and $C[f]$ is an operator on the distribution function $f(J,\theta,t)$ that describes the effect of small-scale kicks (`C' for `collision'). 

In terms of the canonical variables $(q,p)$, the model for small-scale kicks in \S\ref{sec:small} implies that 
\begin{equation}
    C[f]=\half D_\mathrm{sm}\left(\frac{\p^2f}{\p q^2} + \frac{\p^2f}{\p p^2}\right);
\end{equation}
this is a special case of the Fokker--Planck equation. Converting to action-angle variables using equations (\ref{eq:pqjt}) we have 
\begin{equation}
    C[f]=D_\mathrm{sm}\left(\frac{\p}{\p J}J\frac{\p f}{\p J}+ \tfrac{1}{4}\frac{\p^2f}{\p \theta^2}\right).
\end{equation}

We are interested in the late-time behavior of the snail. In this case the spiral is tightly wound, so derivatives with respect to $J$ are the largest terms in $C[f]$. Near a reference action $J_0$ we can write $J=J_0+\Delta J$ and approximate
\begin{equation}
    C[f]=D_\mathrm{sm} J_0\frac{\p^2 f}{\p (\Delta J)^2}.
\end{equation}
Similarly we can approximate $\Omega(J)=\Omega_0+\Omega_1\Delta J$ where $\Omega_0\equiv\Omega(J_0)$ and $\Omega_1=\mathrm{d}\Omega/\mathrm{d}J|_{J_0}$. Then the Boltzmann equation becomes
\begin{equation}
    \frac{\p f}{\p t} + (\Omega_0+\Omega_1\Delta J)\frac{\p f}{\p\theta}=D_\mathrm{sm}J_0\frac{\p^2 f}{\p (\Delta J)^2}.
\end{equation}
The general solution of this partial differential equation can be written in terms of Airy functions, but for our purposes it is sufficient to examine the particular solution 
\begin{align}
f(J_0+\Delta J,&\theta,t)\nonumber \\ &=A\cos\left\{m[\theta - (\Omega_0+\Omega_1\Delta J)(t-t_i)-\theta_i]\right\}\nonumber \\
&\quad\times \exp[-\tfrac{1}{3}m^2D_\mathrm{sm}J_0\Omega_1^2(t-t_i)^3].
\end{align}
where $A$ is an arbitrary constant. The cosine term represents the usual undamped snail (cf.\ eq.\ \ref{eq:wrap}). The exponential shows that the amplitude of the snail is damped by the factor 
\begin{equation}
    F(\tau)=\exp(-\tau^3/t_\mathrm{max}^3),
    \label{eq:dampdef}
\end{equation}
where $\tau$ is the age of the snail and 
\begin{equation}
    t_\mathrm{max}=\left(\frac{3}{m^2D_\mathrm{sm}J_0\Omega_1^2}\right)^{1/3}.
\end{equation}

We may specialize to $m=1$ (one-armed spiral), rewrite $\Omega_1$ in terms of the shear $\Gamma(J)$ (eq.\ \ref{eq:fdef}), and replace $D_\mathrm{sm}$ by $(1-g)\langle J\rangle/T$ where $T$ is the age of the Galaxy and $g$ is the fraction of the disc thickness due to large-scale kicks. Then
\begin{equation}
    t_\mathrm{max}=\left[\frac{3J_0P^2(J_0)T}{(2\pi)^2(1-g)\langle J\rangle \Gamma^2(J_0)}\right]^{1/3}.
    \label{eq:tmaxdef}
\end{equation}
If the reference action is set to the mean action, $J_0=\langle J\rangle$, and we assume that $g\ll1$ then
\begin{equation}
   t_\mathrm{max}=0.424\, \Gamma^{-2/3}P^{2/3}T^{1/3},
   \label{eq:tmaxmean}
\end{equation}
close to the result derived by approximate arguments in \S\ref{sec:snaildyn}.

Notice that the damping is quite fast, once it begins -- faster than an exponential or even a Gaussian. Over a factor of two in age, from $\tau=2^{-1/2}t_\mathrm{max}$ to $\tau=2^{1/2}t_\mathrm{max}$, the snail damps from 70\% to 6\% of its initial amplitude. 

\bsp

\label{lastpage}

\end{document}